\documentclass[useAMS,usenatbib]{aa}
\usepackage{txfonts,color}
\usepackage[authoryear]{natbib}
\bibpunct{(}{)}{;}{a}{}{,}
\usepackage{supertabular}
\usepackage[dvips]{graphicx}
\usepackage{enumerate}

\newcommand{\simgt}{\lower.5ex\hbox{$\; \buildrel > \over \sim \;$}}
\newcommand{\simlt}{\lower.5ex\hbox{$\; \buildrel < \over \sim \;$}}

\def\farcs{\hbox{$.\!\!^{\prime\prime}$}}

\def\ls{\mathrel{\hbox{\rlap{\hbox{\lower4pt\hbox{$\sim$}}}\hbox{$<$}}}}
\def\gs{\mathrel{\hbox{\rlap{\hbox{\lower4pt\hbox{$\sim$}}}\hbox{$>$}}}}

\def\V606{\mathrel{V_{606}}}
\def\Planck{\emph{Planck$\;$}}

\def\XMM{\emph{XMM-Newton}$\;$}
\def\ROSAT{\emph{ROSAT$\;$}}
\def\Euclid{\emph{Euclid$\;$}}
\def\eROSITA{\emph{eROSITA$\;$}}

\begin{document}

\title{HIFLUGCS: X-ray luminosity -- dynamical mass relation and
  its implications for mass calibrations with the SPIDERS and
  4MOST surveys}

\author{Yu-Ying Zhang\inst{1},
Thomas H. Reiprich\inst{1},
Peter Schneider\inst{1},
Nicolas Clerc\inst{2},
Andrea Merloni\inst{2},
Axel Schwope\inst{3},
Katharina Borm\inst{1},
Heinz Andernach\inst{4},
C\'esar A. Caretta\inst{4,5},
\and
Xiang-Ping Wu\inst{6} 
}
\institute{Argelander-Institut f\"ur Astronomie, Universit\"at Bonn, Auf dem
  H\"ugel 71, 53121 Bonn, Germany 
\and Max-Planck-Institut f\"ur extraterrestrische Physik, Giessenbachstra{\ss}e, 85748 Garching, Germany
\and Leibniz-Institut f\"ur Astrophysik Potsdam (AIP), An der Sternwarte 16, 14482 Potsdam, Germany 
\and Departamento de Astronom\'{\i}a, Universidad de Guanajuato, AP 144, Guanajuato CP 36000, Mexico
\and Laboratoire d'Astrophysique de Marseille, Aix-Marseille Universit\'{e},
   38 rue Frederic Juliot-Curie, 13388 Marseille, France
\and National Astronomical Observatories, Chinese Academy of Sciences, Beijing 100012, China}

\authorrunning{Zhang et al.}

\titlerunning{HIFLUGCS: $L-M$ relation and its implications for mass calibrations with the SPIDERS and 4MOST surveys}

\date{Received 20/05/2016 / Accepted 23/08/2016}

\offprints{Y.-Y. Zhang}

\abstract{We present the X-ray luminosity versus dynamical mass
  relation for 63 nearby clusters of galaxies in a flux-limited
  sample, the HIFLUGCS (consisting of 64 clusters). The luminosity
  measurements are obtained based on $\sim 1.3$~Ms of clean \XMM data
  and \ROSAT pointed observations. The masses are estimated using
  optical spectroscopic redshifts of 13647 cluster galaxies in
  total. We classify clusters into disturbed and
    undisturbed ones, based on a combination of the X-ray luminosity
    concentration and the offset between the brightest cluster galaxy and
    X-ray flux-weighted center. Given sufficient numbers
  (i.e. $\ge$45) of member galaxies in computing the dynamical masses,
  the luminosity versus mass relations agree between the disturbed and
  undisturbed clusters. The cool-core clusters still dominate the
  scatter in the luminosity versus mass relation even when a core
  corrected X-ray luminosity is used, which indicates that the scatter
  of this scaling relation mainly reflects the structure formation
  history of the clusters. As shown by the clusters with a small
  number of spectroscopically confirmed members, the dynamical masses
  can be underestimated and thus lead to a biased scaling relation. To
  investigate the potential of spectroscopic surveys to follow up
  high-redshift galaxy clusters/groups observed in X-ray surveys for
  the identifications and mass calibrations, we carried out
  Monte-Carlo re-sampling of the cluster galaxy redshifts and
  calibrated the uncertainties of the redshift and dynamical mass
  estimates when only reduced numbers of galaxy redshifts per cluster
  are available. The re-sampling considers the SPIDERS and 4MOST
  configurations, designed for the follow-up of the \eROSITA clusters,
  and was carried out for each cluster in the sample at the actual
  cluster redshift as well as at the assigned input cluster redshifts
  of 0.2, 0.4, 0.6, and 0.8. For following up very
    distant cluster/groups, we also carried out the mass calibration
    based on the re-sampling with only ten redshifts per cluster, and
    redshift calibration based on the re-sampling with only five and
    ten redshifts per cluster, respectively. Our results demonstrate
  the power of combining upcoming X-ray and optical spectroscopic
  surveys for mass calibration of clusters. The scatter in the
  dynamical mass estimates for the clusters with at least ten members
  is within 50\%.
}

\maketitle

\section{Introduction}

Galaxy clusters represent the place where astrophysics and cosmology
meet: while their overall internal dynamics is dominated by gravity,
the astrophysical processes taking place on galactic scales leave
observable imprints on the diffuse hot gas trapped within their
potential wells (Giacconi et al. 2009). Galaxy clusters
(e.g. Vikhlinin et al. 2009a, 2009b; Mantz et al. 2010), in
combination with supernovae (e.g. Riess et al. 2011), cosmic microwave
background (e.g. Bennett et al. 2013; Hinshaw et al. 2013; \Planck
Collaboration et al. 2015a), baryon acoustic oscillations (e.g. Dawson
et al. 2013) and cosmological weak lensing (e.g. Schrabback et
al. 2010; Laureijs et al. 2011; Marian et al. 2011; Heymans et
al. 2013), can constrain the dark energy equation-of-state parameter,
$w$, at both late and early cosmological epochs. Upcoming experiments
such as \eROSITA will improve the statistical power of cluster
cosmology by a few orders of magnitude (e.g. Predehl et al. 2010;
Merloni et al. 2012). However, there are two major
challenges in upcoming surveys for high-precision cluster cosmology:
accurate mass calibrations and efficient follow-up,
such as optical identifications and redshift
measurements.

The cluster cosmological applications are degenerate with the mass
calibrations of galaxy clusters (e.g. Zhang \& Wu 2003; Stanek et al.
2010; Merloni et al. 2012). Large X-ray surveys select galaxy clusters
by their observables (e.g. Ebeling et al. 2000; B\"ohringer et al.
2004; Clerc et al. 2012; Hilton et al. 2012; Takey et al. 2013),
particularly the X-ray luminosity ($L$), rather than by their masses
($M$). The $L-M$ relation is required to recover the selection
function in terms of cluster masses and predict the cluster masses,
hence the cluster mass function. The bolometric X-ray luminosity
versus mass relations calibrated with different samples differ
significantly (e.g. Chen et al. 2007; Maughan 2007; Pacaud et al.
2007; Pratt et al. 2009; Mantz et al. 2010; Reichert et al. 2011), in
which the slope varies from $\sim$1.5 to $\sim$2.0. The predicted
numbers of detected clusters in the \eROSITA survey based on different
calibrations of the $L-M$ scaling relations differ by up to a factor
of two (e.g. Pillepich et al. 2012). Apart from the luminosity
segregation between cool-core (CC) and non-cool-core (NCC) clusters,
merging affects the properties of both the intracluster medium (ICM)
and cluster galaxies (e.g. Poole et al. 2006; Evrard et al. 2008), but
on different time scales (e.g. Roettiger et al. 1997). Mass estimates
from e.g. optical spectroscopy and gravitational lensing, independent
of the X-ray luminosity measurement, provide an X-ray blind reference
of the cluster mass to calibrate the $L-M$ relation (e.g. Kellogg et
al. 1990; Wu et al. 1998, 1999; Zhang et al. 2008; Leauthaud et al.
2010; Israel et al. 2014, 2015; von der Linden et al. 2014). However,
the dynamical mass estimates are sensitive to the cluster galaxy
selection and member statistic, and overestimate the mass of a merging
cluster when the merger axis is along the line of sight (e.g. Biviano
et al. 2006; Gifford \& Miller 2013; Old et al. 2013; Saro et al.
2013; Wu \& Huterer 2013; Rabitz et al. submitted).

The HIghest X-ray FLUx Galaxy Cluster Sample (HIFLUGCS, Reiprich \&
B\"ohringer 2002) is a flux-limited sample of 64 clusters selected
from the \ROSAT All-Sky Survey (RASS; Ebeling et al. 2000; B\"ohringer
et al. 2004). We analyze high-quality \XMM and \ROSAT pointed
observations as well as optical spectroscopic redshifts of 13650
cluster galaxies for all 64 HIFLUGCS clusters. Excluding 2A~0335+096
with only three redshifts, we present the X-ray luminosity versus
dynamical mass relation for the 63 HIFLUGCS clusters, and quantify the
impact of mergers as well as CC systems on the $L-M$ relation. A
simultaneous mass calibration and cosmological application procedure
(e.g. Bocquet et al. 2015; Mantz et al. 2015) can break the degeneracy
between them. Such an approach is promising for the use of the
combined X-ray and optical surveys in the near future because a vast
number of telescopes are dedicated to carry out the optical
spectroscopic surveys of galaxy clusters,
e.g. eBOSS/SPIDERS\footnote{www.sdss.org/surveys/eboss} (Clerc et
al. 2016)
and 4MOST\footnote{www.4most.eu}.
This method, however, relies
not only on independent measurements of at least two cluster
quantities, e.g. X-ray luminosity and dynamical mass, but also on well
understood knowledge on the uncertainties and potential biases in both
measured quantities (e.g. Applegate et al. 2014; von der Linden et
al. 2014; Planck collaboration 2015b).

Furthermore, efficient follow-up identifications of a large number of
galaxy clusters become one of the main tasks in upcoming surveys for
high-precision cluster cosmology (e.g. Merloni et al. 2012; Nandra et
al. 2013; Pointecouteau et al. 2013). For optimizing the optical
spectroscopy follow-up, it is invaluable to investigate in which
redshift range existing and upcoming multi-wavelength surveys are
suitable to identify groups and clusters of galaxies and to measure
their redshifts. For further applications of the optical spectroscopic
follow-up, it is worth testing in which redshift range those
multi-wavelength surveys can provide accurate dynamical masses that
are sufficient for the $L-M$ calibration. 

In this paper, we simulate the optical spectroscopic follow-up of
clusters by Monte-Carlo (MC) re-sampling of the HIFLUGCS cluster
galaxy redshifts in hand according to eight optical spectroscopic
setups. We calibrate the redshift and dynamical mass estimates, and
quantify in which redshift range the tested optical spectroscopic
setups are reliable for measuring the cluster redshifts and dynamical
masses. We organize the paper as follows. In Sect.~\ref{s:ana} we briefly
describe the sample, data and analysis method. The results on the
$L-M$ relation of the HIFLUGCS sample are presented in
Sect.~\ref{s:lm}. The results on the redshift and dynamical mass
calibrations using simulations of the re-sampled cluster galaxy
redshifts are presented in Sect.~\ref{s:sim}. We summarize our
conclusions in
Sect.~\ref{s:con}. Throughout the paper, we assume $\Omega_{\rm m}=0.3$,
$\Omega_\Lambda=0.7$, $\sigma_8=0.81$, and
$H_0=70$~km\,s$^{-1}$\,Mpc$^{-1}$. We quote 68\% confidence
level. Unless explicitly stated otherwise, we apply the BCES bisector
regression method (Akritas \& Bershady 1996)
taking into account measurement errors on both variables to determine
the parameters and their errors in the fit.

\section{Sample, data and analysis method}
\label{s:ana}

The HIFLUGCS is a sample of galaxy clusters with an X-ray flux
limit (0.1--2.4~keV) of $2\times 10^{-11}$ ~erg/s/cm$^{2}$ and
Galactic latitude $|b|>20.0$~degrees covering two thirds of the
sky (Reiprich \& B\"ohringer 2002; Hudson et al.
2010), 
selected from the \ROSAT All-Sky Survey (RASS; Ebeling et al. 2000;
B\"ohringer et al. 2004).

\subsection{X-ray data and analysis}
\label{s:xray}

We analyze nearly $4$~Ms raw data from \XMM for
63 clusters (Zhang et
al. 2009, 2011, 2012). After cleaning and selecting the longest
observation closest to the cluster center, we obtain $\sim 1.3$~Ms
of \XMM data for 59 clusters. We measure the X-ray properties for all 64
clusters in the HIFLUGCS combining
\XMM and \ROSAT data. The X-ray flux-weighted
centroids are listed in Zhang et al. (2011, 2012) with the method
described in Sect.~2.3 of Zhang et al. (2010).

\subsubsection{X-ray luminosity}

The X-ray luminosity is measured within the cluster radius,
$r_{500}$\footnote{This radius is defined as that within which the
  matter over-density is 500 times the critical density of the
  Universe.}, derived from the dynamical mass in Sect.~\ref{s:sigma}
centered on the X-ray flux-weighted centroid except for 2A~0335+096,
which has three galaxy redshifts in total and of which the cluster
radius $r_{500}$ is derived from the mass versus gas mass scaling
relation. We note that in Zhang et al. (2011) the X-ray luminosity was
measured within the $r_{500}$ derived from the gas mass using the
total mass versus gas mass scaling relation in Pratt et al. (2009). To
suppress the scatter of the $L-M$ relation caused by cool cores, we
use the core-corrected X-ray luminosity ($L^{\rm cor}$ hereafter),
which is derived from the integration of the surface brightness
assuming a constant value of the surface brightness within
$0.2r_{500}$ equal to the value at $0.2 r_{500}$, $S_{\rm X}(R <
0.2r_{500})=S_{\rm X}(0.2r_{500})$, following Zhang et
al. (2007). Here $R$ is the projected cluster-centric distance. We
note that this correction is only applied in determining the X-ray
luminosity. We use the core-corrected bolometric X-ray luminosity
(Table~\ref{t:log}) for the luminosity versus mass relation of the
HIFLUGCS throughout the paper.

Some past work used a radius larger than $r_{500}$
(e.g. Reiprich \& B\"ohringer 2002) to derive the X-ray luminosity,
and some used different methods to account for the cool cores
(e.g. Pratt et al. 2009). Different methods in deriving the X-ray
luminosity may lead to different normalization values of the $L-M$
relation, and also to different slopes. We therefore tabulate the
ratios of the X-ray luminosity derived within different annuli for the
whole HIFLUGCS of 64 clusters in Table~\ref{t:lbratio}. Those values
help to ensure a fair comparison between our results and those in
other papers. Since the X-ray luminosity varies very little
with the truncation radius, the correlation introduced by using the
dynamical mass-determined $r_{500}$ in deriving the X-ray luminosity
to the $L^{\rm cor}-M$ relation is negligible.


\subsection{Optical data and analysis}
\label{s:opt}

The positions of the brightest cluster galaxies (BCGs) for 63 clusters
are listed in Table~1 in Zhang et al. (2011) and for
RXC~J1504.1$-$0248 in Zhang et al. (2012). We obtain the velocity of
the cluster galaxies from the literature (updated until April 2013,
including the compilation in Andernach et al. 2005 and Zhang et
al. 2011, 2012). Since the individual error estimates are
inhomogeneous, we decided not to weight the calculation of the average
velocity. When there is more than one velocity per galaxy available,
we calculate an average of the measurements, excluding discordant
values and those with large errors.

We checked recent redshift surveys, and found significant ($>90$)
numbers of new redshifts for Abell clusters: A2029 from Tyler et
al. (2013), A2142 from Owers et al. (2011), A2255 from Tyler et
al. (2014), and A2256 from Rines et al. (2016). For 2A~0335+96, NED
now offers more redshifts from Huchra et al. (2012), resulting in 14
galaxies between 8400 and 13200~km/s, with a velocity dispersion of
about 720~km/s. For NGC~1550, we find 42 galaxies with a velocity
dispersion of about 700~km/s.

This work focuses on calibrating the uncertainties of the redshift and
dynamical mass estimates when only reduced numbers of galaxy redshifts
per cluster are used. The impact on our study by adding these new
redshifts mentioned above is negligible. We thus decide to stay with
our compilation without update. In our upcoming study, we aim to
follow up those clusters with low numbers of spectroscopic members to
ensure more than about 45 members for all clusters in the sample in
order to derive robust mass estimates for all individual clusters, and
calibrate the luminosity versus mass relation taking into account the
sample selection effect, i.e. the Malmquist bias (Malmquist 1922) that
arises from working with a flux-limited sample. Therefore, we will
revise the dynamical mass estimates including these above new
redshifts therein.

\subsubsection{Cluster galaxy selection}
\label{s:membcg}

In hierarchical structure-formation scenarios, the spherical infall
model predicts a trumpet-shaped region in the diagram of line-of-sight
velocity versus projected distance, the so-called caustic (e.g.
Kaiser 1987). The boundary of the caustic defines galaxies inside as
cluster galaxies and those outside as fore- and background
galaxies. For each cluster, we plot the line-of-sight velocity of the
selected galaxies as a function of their projected distance from the
BCG, and locate the caustic which efficiently excludes interlopers
(e.g. Diaferio 1999; Katgert et al. 2004; Rines \& Diaferio 2006). We
consider only the galaxies inside the caustic as cluster galaxies, and
exclude the others from the subsequent analysis. We list the number of
cluster galaxies ($n_{\rm gal}$) for the HIFLUGCS in
Table~\ref{t:log}. We gathered redshifts of a total of 13650 cluster
galaxies after the caustic member selection. Since 2A~0355+096 has
only three redshifts, we excluded it from the study involving cluster
redshift and dynamical mass estimates. The median number of
spectroscopic members for the remaining sample is 188 per cluster.

There are another six clusters that have fewer than 45 cluster
galaxies with spectroscopic redshifts in the HIFLUGCS (i.e. 13
galaxies for A0478, 22 for NGC1550, 42 for EXO~0422$-$086, 37 for
Hydra~A, 20 for S1101, and 44 for A2597), which were maintained in the
redshift and dynamical mass study. All six systems with $<45$ cluster
galaxy redshifts are CC clusters. The reason for having $<45$ cluster
galaxy redshifts is that these clusters are not covered by large
spectroscopic surveys such as SDSS or 2dF/6dF. We aim to observe them
with ground-based telescopes. Recently, we obtained new redshifts from
VLT/VIMOS for Abell~S1101 (Rabitz et al. submitted).

\subsubsection{Redshift, dynamical mass and cluster radius}
\label{s:sigma}

We apply the bi-weight estimator (e.g. Beers et al. 1990) to the
cluster galaxies to measure the cluster redshift ($z$) and velocity
dispersion ($\sigma$). The errors are estimated through 1000 bootstrap
simulations. 

The caustic method mentioned in Sect.~\ref{s:membcg} can provide an
estimate of the escape velocity and thus the matter distribution for
the halo according to the distribution of the caustic amplitude
(e.g. Diaferio 1999). However, only with a large number of
spectroscopic members, e.g. $\sim$200 members, one is likely to
recover a rather complete sampling of the caustic, which yields a
robust measurement of the underlying dark matter distribution of the
cluster as shown by the simulations in e.g. Serra \& Diaferio
(2013). Their simulations also show that a lower number of members
thus tends to provide a slightly reduced amplitude of the caustic due
to under-sampling of the caustic, which causes underestimation of the
total mass.

To optimize optical spectroscopic follow-up for galaxy clusters,
particularly at high redshifts where the cluster galaxy member
statistic is low, we focus on the method to derive the dynamical mass
based on a small number of cluster galaxies, differing from the
caustic method. There are a number of such methods calibrated by
samples of simulated clusters (e.g. Biviano et al. 2006; Munari et
al. 2013). We note that the difference between the dynamical
masses derived with those methods is rather small which is within the
uncertainties of our dynamical mass estimates. Apart from this, we
focus only on the relative variation in the dynamical mass
measurements that is normalized to the input dynamical mass in the
investigations of the systematic bias and uncertainties of redshift
and dynamical mass estimates based on MC re-sampling
(Sect.~\ref{s:sim}).

Our method of mass estimation is based entirely on the velocity
dispersion estimate as explained in Sect.~3 of Biviano et al. (2006,
$M_{\sigma}$ therein). We follow the Navarro, Frenk \& White (NFW,
1997) model used in Biviano et al. (2006) to derive $M_{500}$ from
$M_{\sigma}$. The dimensionless Hubble parameter,
$E(z)=\sqrt{\Omega_{\rm m}(1+z)^3
  +\Omega_{\Lambda}}$, is added to their formulation since not all our
clusters are at $z\sim 0$.
%
%
In this work, we use the dynamical mass at the radius within which
the over-density is 500 times the critical density as the
cluster mass ($M_{500}$, Table~\ref{t:log}). We derive the cluster
radius ($r_{500}$) from the dynamical mass ($M_{500}$),
different from the method used by Zhang et al. (2011).

\subsection{Cluster morphology}
\label{s:mor}

Although the cluster properties are driven by gravity, predominantly
from dark matter, the baryon physics on small scales modifies the
scaling relation from the self-similar prediction through cooling,
merging and feedback from star formation and AGN activities. We
divided the HIFLUGCS into undisturbed versus disturbed subsamples as
well as CC and NCC subsamples in order to study the impact of mergers
and CC clusters, respectively, on our results.

\subsubsection{Undisturbed versus disturbed clusters}

Many studies found that a large offset between the BCG and X-ray
peak/centroid likely indicates that more merging has
happened in the past (e.g. Katayama et al. 2003; Hudson et al. 2010;
Zhang et al. 2010, 2011). We convert the angular separation between
the X-ray flux-weighted center and BCG position into the physical
separation at the cluster redshift
for all 64 clusters in the HIFLUGCS. As shown in Zhang et al. (2011),
the offset in projection between the X-ray flux-weighted center and
BCG position scaled by the cluster radius, ${\Delta}R/r_{500}$, serves
rather well to separate the disturbed from the undisturbed clusters.

The luminosity concentration is also one of the parameters suitable to
separate the clusters (e.g. Santos et al. 2008). Here we set this
parameter to be $c_{\rm L}=L(R\le r_{500})/L(R\le 0.2r_{500})$, and we
integrate the surface brightness within a projected cluster-centric
distance of $r_{500}$ and $0.2 r_{500}$ to derive the bolometric X-ray
luminosity of $L(R\le r_{500})$ and $L(R\le 0.2r_{500})$,
respectively. All HIFLUGCS clusters have also \emph{Chandra} data, but
mainly in the field of the cluster cores, not even out to $0.2r_{500}$
for a number of cases. We thus prefer to rely on \XMM data to compute
the $c_{\rm L}$ parameter.

\begin{figure*}
\begin{center}
\hspace{-1cm}
\includegraphics[angle=90,width=6.8cm]
{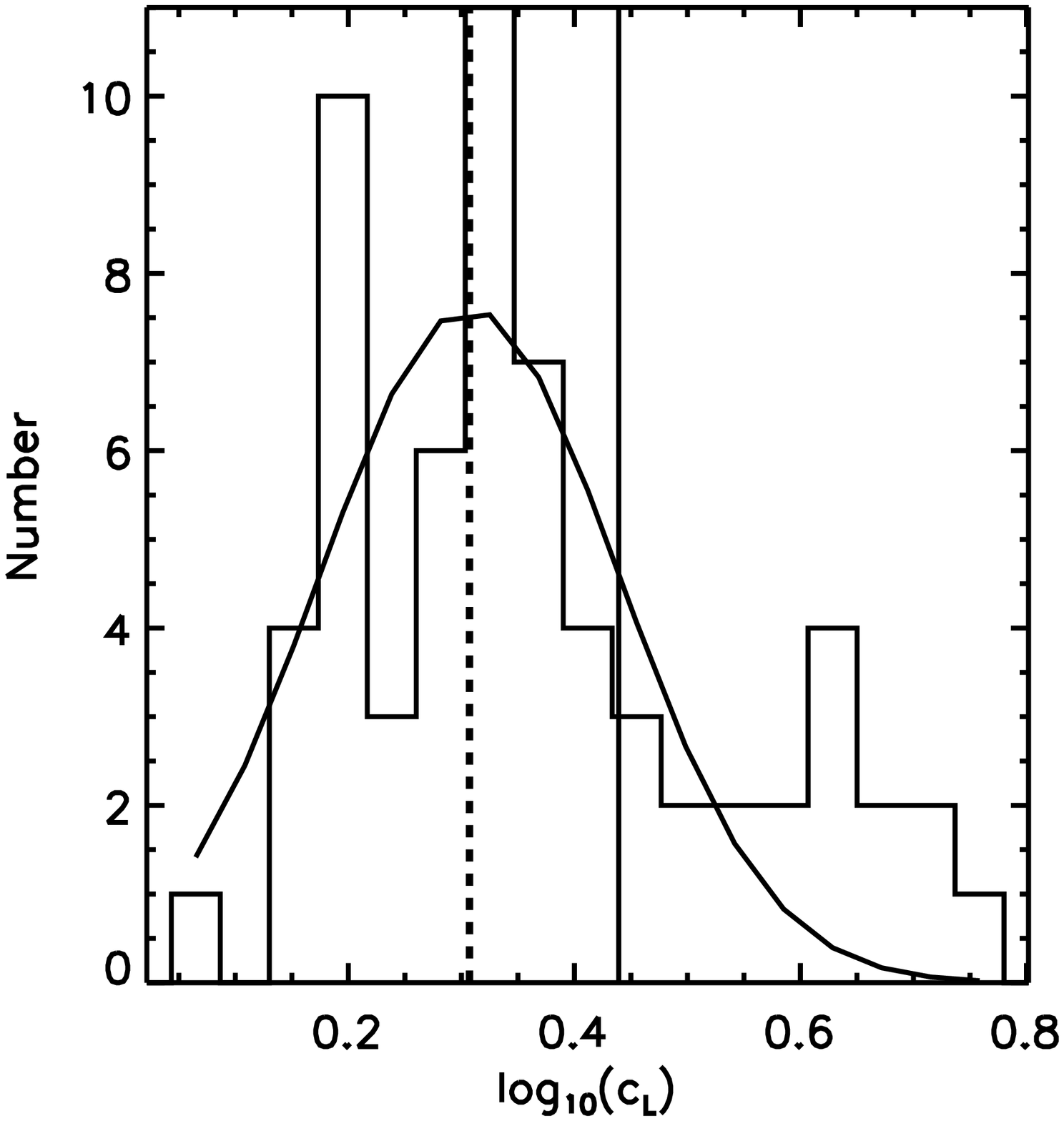}
\hspace{-1cm}
\includegraphics[angle=90,width=6.8cm]
{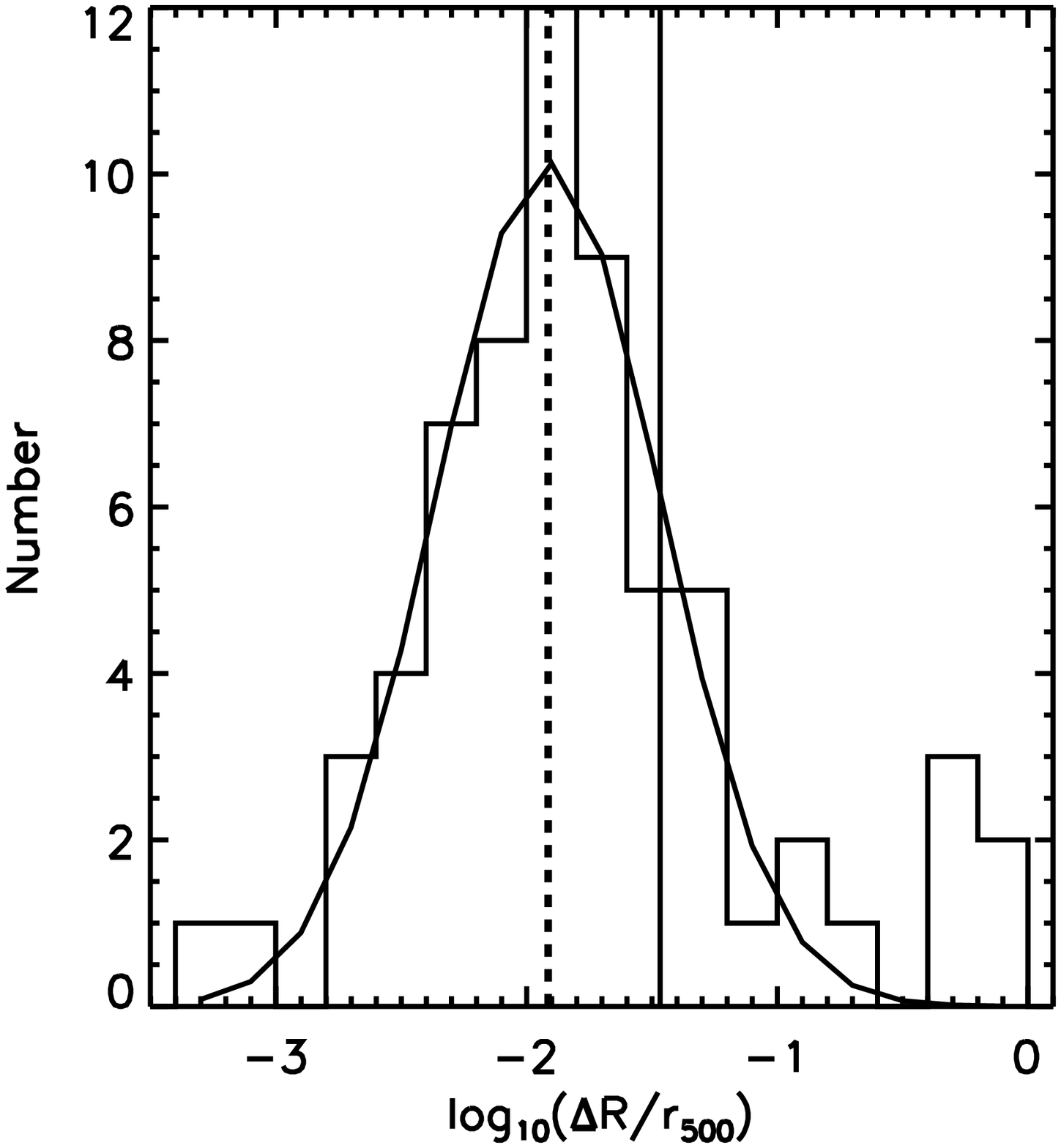}
\hspace{-1cm}
\includegraphics[angle=90,width=6.8cm]
{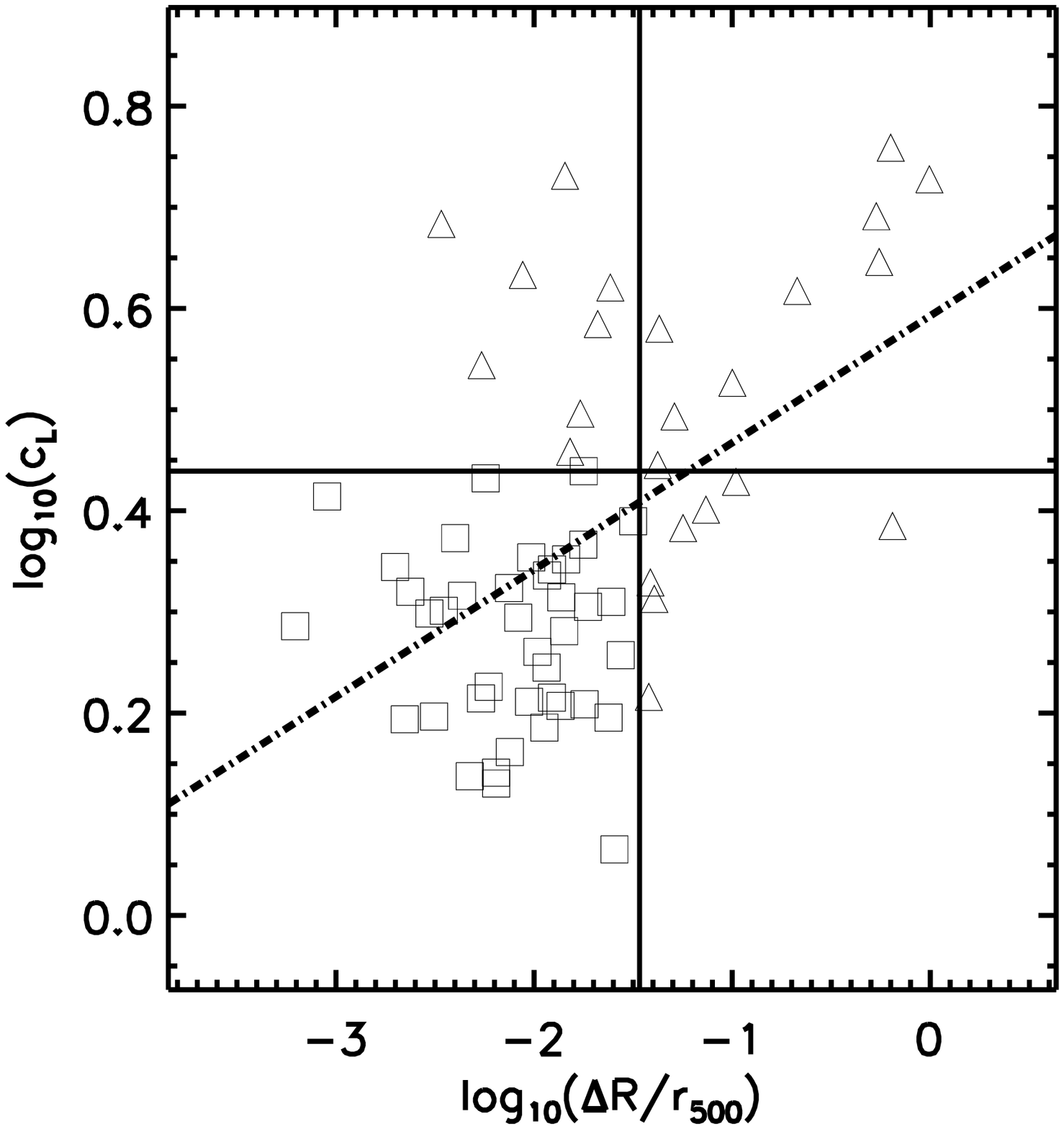}
\hspace{-1cm}
\end{center}
\caption{{\it Left and middle panels:} Histograms of the luminosity
  concentration (left) and the offset between the BCG and X-ray
  flux-weighted centroid (middle) of all 64 HIFLUGCS clusters with
  their Gaussian fits shown as solid curves. The mean of the Gaussian
  fits and their 1-$\sigma$ clipping are shown as dashed and solid
  vertical lines, respectively. {\it Right panel:} $c_{\rm L}$ versus
  ${\Delta}R/r_{500}$ with its best fit, $\log_{10}c_{\rm L}=(0.593\pm
  0.053)+(0.125\pm 0.028)\log_{10}({\Delta}R/r_{500})$, as dash-dotted
  line. The vertical and horizontal solid lines denote the 1-$\sigma$
  clipping, below which the clusters are considered as undisturbed
  ones (open squares). The disturbed clusters are shown as open
  triangles.}
\label{f:histcloff}
\end{figure*}

In the following we combine the cuts of $c_{\rm L}$ and
${\Delta}R/r_{500}$ to effectively separate the sample into
undisturbed and disturbed clusters. As shown in Fig.~\ref{f:histcloff}
and Table~\ref{t:histcloff}, the histograms of both $\log_{10}c_{\rm
  L}$ and $\log_{10}({\Delta}R/r_{500})$ of the 64 clusters follow
approximately Gaussian distributions. The best power-law fit between
the two parameters for the 64 clusters is $\log_{10}c_{\rm
  L}=(0.593\pm 0.053)+(0.125\pm 0.028)\log_{10}({\Delta}R/r_{500})$.
We divide the whole sample into disturbed and undisturbed clusters
using the 1$\sigma$-clipping of the Gaussian-mean values of these two
histograms. Those with low values of both $c_{\rm L}$ and
${\Delta}R/r_{500}$ are considered as undisturbed clusters (see the
right panel of Fig.~\ref{f:histcloff}). The cuts divide the
  sample into 39 undisturbed clusters and 24 disturbed clusters.

\subsubsection{CC versus NCC clusters}

The gas mass versus total mass relation in recent work (e.g. Arnaud et
al. 2005; Zhang et al. 2008) shows a slope shallower than unity,
the self-similar prediction, which may be accounted for by the mass
dependence of the gas mass fraction in CC and NCC systems (e.g. Eckert
et al. 2012). Therefore, we also investigate our results for the
subsamples of CC and NCC clusters, respectively.

The central cooling time can be accurately estimated from
\emph{Chandra} data because of its smaller point-spread function (PSF). We 
used the central cooling time calculated at $0.004 r_{500}$ from
Eq.~(15) in Hudson et al. (2010), and divided the sample of
the 64 clusters into 28 CC clusters and 36 NCC
clusters as listed in Table~2 in Zhang et al. (2011).


\section{X-ray luminosity versus dynamical mass} 
\label{s:lm}


We carried out the X-ray and optical analyses independently, apart
from taking the dynamical mass-determined $r_{500}$ in computing the
X-ray bolometric luminosity. Since the luminosity values measured
within $r_{500}$ and 2.5$r_{500}$ differ only by $\sim$15\% on average
as shown in Table~\ref{t:lbratio}, using the cluster radius, $r_{500}$,
derived from the dynamical mass, in computing the X-ray bolometric
luminosity shall not cause any significant bias in our result. 

We fit the $L^{\rm cor}-M$ relation by a power-law,
\begin{equation}
\log_{10}\widetilde{L}=A\log_{10}\widetilde{M}+B~,
\label{e:lm}
\end{equation}
in which $\widetilde{L}=\frac{L^{\rm cor}}{E(z)10^{44}{\rm erg/s}}$
and $\widetilde{M}=\frac{M\,E(z)}{10^{14}M_{\odot}}$. Note that the
HIFLUGCS is a flux-limited sample, which consists of clusters covering
a broad range of redshifts (i.e. from $z=0.0037$ to $z=0.2153$).
Therefore, we include the $E(z)$ factor in the scaling relation to
account for the evolution of the geometry of the Universe in this
  redshift range. We assume a constant standard deviation of the
luminosity from this relation in logarithmic scale, and use
$\sigma_{\log_{10}L}$ for the intrinsic scatter of the luminosity in
logarithmic scale. We choose the relatively simple algorithm given in
Akritas \& Bershady (1996) to perform the fit to the relation
$y=Ax+B$, in which
\begin{equation}
  \chi^2= \sum_{i} \frac{(y_i - A x_i -B)^2}{\sigma^2_{y,i}+A^2 \sigma^2_{x,i}} ~.
\label{e:chi2}
\end{equation}
The scatter of $y$ consists of two components, the statistical and
intrinsic scatter, which satisfy $\sigma^2_{y,i}=\sigma^2_{y,i,{\rm
sta}}+\sigma^2_{y,i, {\rm int}}$ (e.g. Weiner et al.  2006). In
Table~\ref{t:lm}, we list the best-fit parameters also including the
normalization values setting the slope to the self-similar prediction,
namely 4/3. Note that the data are still in agreement
with a slope of 4/3 within $2\sigma$ uncertainties of the best-fit
slope parameters.

\subsection{$L^{\rm cor}-M$ relation for the $n_{\rm
      gal} \ge 45$ clusters}

The dynamical masses tend to be underestimated for
the clusters with too few redshift measurements according to the
clusters in simulations (e.g. Biviano et al. 2006; Zhang et
al. 2011). The effect becomes negligible (i.e. on a
level of a few per cent) when clusters with a sufficient number of
spectroscopic cluster galaxies are considered.

In Zhang et al. (2011) we found that the velocity dispersion
measurements, and thus the dynamical masses, may be significantly
underestimated for the clusters with fewer than 45 spectroscopic
member galaxies in the sample. To avoid any bias due to extreme
outliers with too few cluster galaxies with spectroscopic redshifts,
we show the $L^{\rm cor}-M$ relations for the 63 clusters
(Fig.~\ref{f:lm12}a) and for the 57 clusters that have at least 45
cluster galaxies with spectroscopic redshifts (Fig.~\ref{f:lm12}b),
respectively. For the 63 clusters, the slope, $1.29\pm0.09$, agrees
with the self-similar prediction. However, most recent work shows
steeper slopes, e.g. $1.77\pm 0.07$ in Pratt et al. (2009). Taking a
close look, the systems with very few redshifts appear to cause the
shallow slope. In contrast to that of the 63 clusters, the slope of
the 57 clusters is steeper. The selection effect, i.e. the Malmquist
bias that should be corrected for a flux-limited sample, should lead
to an even steeper slope for the 57 clusters, which tends to be in
better agreement with recent X-ray calibrated $L-M$ relations
(e.g. Pratt et al. 2009). At the same time, the fact that the six
  discarded clusters are all CC clusters may also have an impact on
the total best-fit slope, although there appears little difference
between the slopes of the $L-M$ relations between CC and NCC clusters.

\subsection{$L^{\rm cor}-M$ relations for the CC/NCC
    and undisturbed/disturbed subsamples}

\begin{figure*}
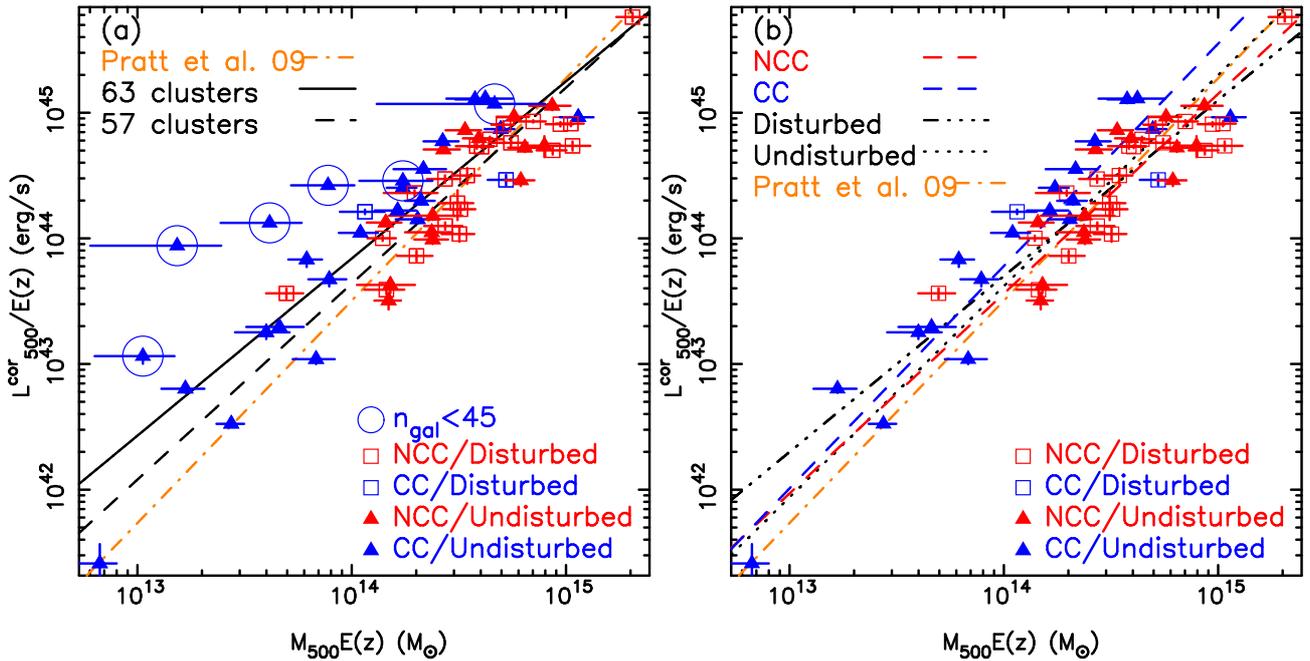

\begin{center}
\includegraphics[angle=270,width=8.5cm]
{plots/lbm0500_conloff_n63_vs_n57_vs_xcali_l.ps}
\includegraphics[angle=270,width=8.5cm]
{plots/lbm0500_conloff_n57_cc2_dis2_l_pratt09.ps}
\end{center}
\caption{(a) Core-corrected bolometric X-ray luminosity versus
  dynamical mass for the 63 clusters with the best power-law fits for
  the 63 and 57 ($n_{\rm gal}\ge 45$) clusters as black solid and
  dashed lines, respectively. (b) Core-corrected X-ray bolometric
  luminosity versus dynamical mass for the 57 ($n_{\rm gal}\ge 45$)
  clusters with the best power-law fits for the NCC, CC, disturbed,
  and undisturbed clusters as red dashed, blue dashed, black
  dot-dot-dot-dashed, and black dotted lines, respectively.}
\label{f:lm12}
\end{figure*}

\begin{figure*}
\begin{center}
\includegraphics[angle=90,width=12cm]{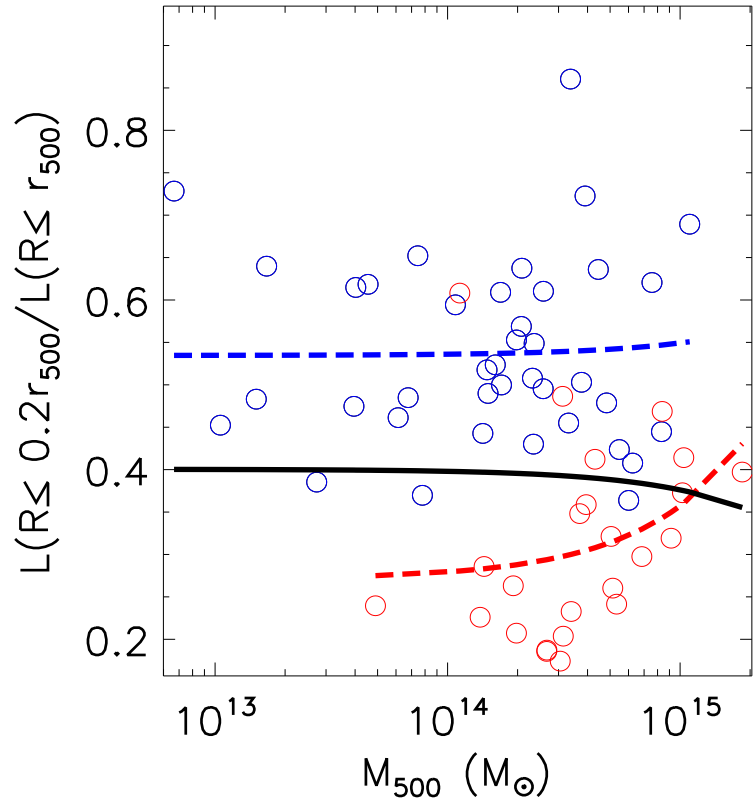}
\end{center}
\caption{X-ray core-to-$r_{500}$ luminosity fraction versus dynamical
  mass for the CC (blue) and NCC (red) clusters. The curves show the
  regression fits for all (black), CC (blue) and NCC (red) clusters,
  respectively. There is no obvious dependence considering the
  observed scatter.}
\label{f:lm57_3_1}
\end{figure*}

We list the best-fit $L^{\rm cor}-M$ relations regarding cluster
classification in Table~\ref{t:lm}. As shown in Fig.~\ref{f:lm12}b and
Fig.~\ref{f:lm57_3_1}, the NCC clusters are under-sampled in the
low-mass regime since the HIFLUGCS is a flux-limited sample. The slope
parameters agree within 1$\sigma$ error between the undisturbed and
disturbed clusters as well as between the CC and NCC clusters. Note
that there is no obvious mass dependence in the scatter of the
core-to-total luminosity ratio (actually $1/c_{\rm L}$) for the CC and
NCC subsamples, respectively, as shown in Fig.~\ref{f:lm57_3_1}. The
$L^{\rm cor}-M$ relation for the disturbed clusters has smaller
intrinsic scatter than that for the undisturbed clusters. Also the
intrinsic scatter of the NCC clusters is smaller than that of the CC
clusters even when the core-corrected luminosity is in use. Note that
this partially depends on how broad the used cut is to classify the
corresponding cluster subsamples. The normalization of the disturbed
(NCC) clusters is in better agreement with the X-ray calibrated
relation (e.g. Zhang et al. 2008, see Table~3 therein; Pratt et
al. 2009). Assuming a fixed slope of 4/3, the normalization agrees
between different cluster types within the uncertainties.

A comparison of the best fits for the undisturbed (CC) clusters in the
samples of the 57 and 63 clusters in Fig.~\ref{f:lm12} shows clearly
that the scatter for the undisturbed (CC) clusters in the sample of 57
clusters with $N_{\rm gal}>45$ is lower than that for the sample of 63
clusters. As shown in Fig.~\ref{f:lm12}b for the 57 clusters, the
normalization on average agrees within 17\% between the undisturbed
and disturbed clusters, and within 32\% between the CC and NCC
clusters. The slopes still agree among the subsamples. As the scatter
is dominated by the CC clusters, the scatter of the $L^{\rm cor}-M$
relation may mainly be due to the integrated effect of the structure
formation history instead of recent extreme merging.

\section{Systematic bias and uncertainties of redshift and dynamical mass
  estimates based on MC re-sampling}
\label{s:sim}

The results from our sample show that the clusters with low numbers of
spectroscopic members tend to have their dynamical masses
underestimated. It is thus of prime importance to quantify any
systematic bias and uncertainties of the dynamical mass estimate and
their implications on mass calibrations using upcoming optical
spectroscopic surveys.

Both large optical spectroscopic surveys and individual pointed
observations will be used to follow up galaxy clusters detected in
upcoming large X-ray surveys, e.g. \emph{eROSITA}. The huge amount of
optical spectroscopic follow-up data could be potentially combined
with the X-ray data for mass calibrations, in particular, in different
redshift ranges. This allows to constrain the evolution of the mass
scaling relations, which becomes important when using high-redshift
clusters for cosmology. In the following, we calculated the
  cluster redshift and estimated the dynamical mass using subsamples
  of member galaxies which we re-sampled according to eight optical
spectroscopic setups, which demonstrates their ability for studying
mass calibrations using the follow-up of the X-ray detected clusters.
We note that the re-sampled galaxies are selected from the catalog of
the caustic-bounded galaxies. Therefore, the probability of
contamination from interlopers is low. The systematic uncertainties
are mainly caused by
statistical fluctuations in the re-sampling.
Our study is also useful to assign the mass
uncertainty in the stacking analysis using cluster samples with a
rather clean member selection but a small
number of members (e.g.  Clerc et al. submitted).

\subsection{Optical spectroscopic survey setups}

The Extended Baryon Oscillation Spectroscopic Survey (eBOSS; e.g.
Schlegel et al. 2011) is a redshift survey covering a wavelength range
from 340~nm to 1060~nm, with a resolution $R=3000-4800$. This survey
targets objects up to redshift 2. With the eBOSS setup, the
SPectroscopic IDentification of ERosita Sources (SPIDERS; e.g. Merloni
et al. 2012) survey has been designed to follow up X-ray selected
active galactic nuclei (AGN) and clusters over $\sim$7500~degree$^{2}$
area. Before \emph{eROSITA} data will be available, SPIDERS is
targeting \emph{ROSAT} and \emph{XMM-Newton} sources (see Clerc et al.
submitted; Dwelly et al. in prep.). In the later years of
operations, SPIDERS will target \eROSITA sources detected in the first
two years of observations (see Merloni et al. 2012). The 4-metre
Multi-Object Spectroscopic Telescope (4MOST, e.g. de Jong et al. 2012)
is designed to obtain $>$1 million redshifts for $>$50,000 clusters in
the Southern sky in the \eROSITA survey. The \Euclid mission can
follow up high-redshift galaxies through their emission lines (e.g.
Laureijs et al. 2011; Pointecouteau et al. 2013). In this study,
however, we will not discuss the optimization of the follow-up of the
\Euclid survey since \Euclid provides slitless spectroscopy, which has
no fiber constraint. \Euclid will rather focus on, for example, how to
improve the situation due to overlapping spectra in densely populated
regions for optimizing its strategy.

In practice, only a number of cluster galaxies per cluster can be
targeted given a fiber spacing constraint for
multi-object spectroscopic surveys\footnote{Not the
  case for slitless spectroscopic surveys like
  \emph{Euclid}.}. Moreover, toward the high-redshift
regime, a bright limiting magnitude cut can strongly limit the number
of spectroscopic members. To disentangle the two effects, we perform
the investigations in two steps: (i) for the setups using the closest
fiber separation alone and (ii) for the setups including constraints
from both the closest fiber separation and limiting magnitude
cuts. SPIDERS uses actually a minimum fiber separation of
65$^{\prime\prime}$. To demonstrate the impact of the fiber
separation, we use a fiber spacing constraint of 55$^{\prime\prime}$
and 65$^{\prime\prime}$, respectively, in our MC simulations for
SPIDERS. Note that the limiting magnitude is actually the fiber
magnitude.
Toward the very high-redshift regime, optical spectroscopy of several
of the bright galaxies already requires high-sensitivity optical
spectrographs, e.g. on the Very Large Telescope (VLT). We also investigate
the setups with a limited number of cluster redshifts. Below are the
eight setups, in which the BCG is always first selected for
targeting.
\begin{enumerate}[(I)]

\item {SPIDERS\_55}: It has a closest fiber separation of
  55$^{\prime\prime}$. Each position can be pointed only once. 

\item {SPIDERS\_65}: It has a closest fiber separation of
  65$^{\prime\prime}$. Each position can be pointed only once. 

\item {4MOST}: It has a closest fiber separation of
  20$^{\prime\prime}$. Each position can be pointed four times. 

\item {SPIDERS\_55m}: It has a closest fiber separation of
  55$^{\prime\prime}$. Each position can be pointed only once. The
  limiting magnitude is $i_{\rm AB}=21$. 

\item {SPIDERS\_65m}: It has a closest fiber separation of
  65$^{\prime\prime}$. Each position can be pointed only once. The
  limiting magnitude is $i_{\rm AB}=21$. 

\item {4MOSTm}: It has a closest fiber separation of
  20$^{\prime\prime}$. Each position can be pointed four times. The
  limiting magnitude is $r_{\rm AB}=22$.

\item {10zs}: It samples the BCG and nine randomly
  selected remaining member galaxies per cluster. 

\item {05zs}: It samples the BCG and four randomly selected
  remaining member galaxies per cluster.

\end{enumerate}


\subsection{Input cluster redshifts, $z_{\rm in}$, in the I--VI
setups}

Follow-up of high-redshift systems is particularly important to
constrain the evolution of mass calibrations that is still poorly
understood. Technically, nearby clusters can tolerate large values of
the closest fiber separation in the optical spectroscopy. The angular
size of a cluster decreases toward high redshift. The number of
spectroscopic members decreases with redshift given a fixed value of
the closest fiber separation. Therefore, we tested the I--VI setups
assuming input cluster redshifts at the HIFLUGCS cluster redshifts
($z$ in Table~\ref{t:log}) as well as placing the clusters at the
redshift values of 0.2, 0.4, 0.6, and 0.8, respectively. 
We note that the relative rest-frame line-of-sight velocity of any
member galaxy to its host cluster is not modified in this process.

\subsection{Limiting magnitude cuts in the IV--VI setups}

We included the limiting magnitude cuts in re-sampling the member
galaxies in the IV--VI setups as follows.

\subsubsection{Shape of the galaxy luminosity function (GLF)}
\label{s:shapeGLF}

We assume that the observed spectroscopy member galaxies are those
at the bright end of the GLF. We adopted the Schechter
function fit of the GLF of the bright galaxies within a projected
cluster-centric distance of 2~Mpc in the $i$- and $r$-bands (Popesso et
al. 2005, Table 2, local background) for the SPIDERS and 4MOST
surveys, respectively. Given the results in Hansen et al. (2009), we
assume that the shape of the GLF of the bright red members does not
change with cluster-centric radius for simplicity.

\subsubsection{Normalization of the GLF}

There are a number of richness versus mass calibrations (e.g. Hansen
et al. 2005, 2009; Johnston et al. 2007; Reyes et al. 2008). We take
the richness versus weak-lensing mass calibration given by Eq.~(10) in
Hansen et al. (2009), in which the masses are measured within the
radius where the mass over-density is 200 relative to the critical density, and
calculate the richness from the cluster dynamical mass for each
HIFLUGCS cluster.

For simplicity, we assume that the red member galaxies brighter than
$(i^{\ast}+1)$ are similar to those brighter than
$(r^{\ast}+1)$. Therefore, the number of the red member galaxies
brighter than $(i^{\ast}+1)$ is similar to the number of the red
members brighter than $(r^{\ast}+1)$. We obtain the richness by
integrating the GLF down to $(i^{\ast}+1)$ magnitude for the SPIDERS
survey and $(r^{\ast}+1)$ magnitude for the 4MOST survey,
respectively. Assuming the richness obtained from the dynamical mass
equal to the richness derived by integrating the GLF, we obtain the
normalization of the GLF.

\subsubsection{Galactic extinction and $K$-corrections} 

We correct apparent magnitudes for Galactic extinction using the maps
of Schlegel et al. (1998). We assume that photometric errors at
bright magnitudes are 0.05. We use typical colors and their color
uncertainties of luminous red galaxies of 
Maraston et al. (2009), and apply $K$-correction and evolutionary
correction using the luminous red galaxy template in {\rm kcorrect} (v4\_2,
Blanton \& Roweis 2007).
 
\subsubsection{Member selection}
\label{s:member_mc}

The maximum number of member galaxies, $n_{\rm max}$, that can be
spectroscopically followed-up is computed by integrating the GLF down
to the limiting magnitude. In Sect.~\ref{s:shapeGLF}, we simply assumed
that the observed spectroscopy member galaxies are those filling in
the bright end of the GLF. Therefore, if the number of cluster members
for the HIFLUGCS clusters ($n_{\rm gal}$ in Table~\ref{t:log}) is
larger than $n_{\rm max}$, the fraction of the $n_{\rm gal}$
spectroscopic members that are brighter than the limiting magnitude is
$f_{\rm obs}=n_{\rm max}/n_{\rm gal}$; otherwise, all existing
spectroscopic members are brighter than the limiting magnitude,
$n_{\rm max}=n_{\rm gal}$.

Note that many cluster members in the observational sample have no
$i$- and $r$-magnitudes from the NED. As a preliminary selection, we
first take the BCG, and then randomly take $f_{\rm obs}n_{\rm gal}-1$
additional spectroscopic members using the MC method. The initially
selected members are filtered further according to the values of the
closest fiber separation.

\subsubsection{Impact of flux loss of the aperture magnitude}
\label{s:fluxloss}

The angular sizes of very bright cluster galaxies in the nearby
Universe (i.e. $z<0.4$) are usually larger than the fiber aperture.
The flux-loss correction may become important in deriving the apparent
magnitudes from the measured flux in the fiber aperture when
considering the limiting magnitude cut. Additionally, the PSF
  affects the measured apparent magnitudes due to convolution of
  galaxy images caused by the seeing. As shown by our model of the
flux loss in targeting galaxies with spectroscopic fibers in
Appendix~\ref{a:fluxloss}, this effect is negligible and skipped in
our further analyses.

\subsection{Redshift and dynamical mass measurements}
\label{s:zmestimator_mc}

To estimate the average of the systematic uncertainties, we carried
out 500 re-sampling runs per cluster per setup per input cluster
redshift ($z_{\rm in}=z, 0.2, 0.4, 0.6, 0.8$ for I--VI setups and
$z_{\rm in}=z$ for VII--VIII setups). In each run, we measured the
cluster redshift, velocity dispersion and dynamical mass following the
procedure in Sect.~\ref{s:sigma} when the number of re-sampled members
reach ten, $n_{{\rm gal},i}\ge 10$. Otherwise ($n_{{\rm gal},i}<10$), we only
compute the cluster redshift, which is the mean of the member galaxy
redshifts, but not the velocity dispersion and dynamical mass. 

The Gaussian mean and bi-weight (see Sect.~\ref{s:sigma}) mean values
of the histograms of the redshift and dynamical mass estimates from
500 runs based on the re-sampled members agree well in all cases apart
from the 10zs setup. The bi-weight mean of the histograms is in
general closer to the input value than the Gaussian mean for the 10zs
setup. Therefore, we use the bi-weight mean for calculating the
results. The number of member redshifts used per cluster per setup per
input cluster redshift, $n_{{\rm gal},MC}$, is computed as the
bi-weight mean of $n_{{\rm gal},i}$ with $i=0,...,499$. As one may
obtain $n_{{\rm gal},i}< 10$ for some re-sampling runs, $n_{{\rm
    gal},MC}$ can be fewer than ten for some cases involving the
dynamical mass estimates, in which the dynamical mass per cluster per
setup per input cluster redshift, $M_{\rm 500,MC}$, is computed as the
bi-weight mean of $M_{500,j}$ with $j=i$ if $n_{{\rm gal},i}\ge
10$. Simulated clusters with less than ten members are excluded in the
mass computation.

\subsection{Summary of results}

 
\subsubsection{Fraction of selected members}

Using the collisional distance constraint alone
(setups I--III), almost all members were selected in
the 4MOST setup, and more than 95\% of the members were selected in
the SPIDERS\_55 and SPIDERS\_65 setups. In the case of ten/five
redshifts per cluster, only a small fraction of the members are
selected. The fraction of selected members including the limiting
magnitude cut (setups IV--VI) decreases faster toward
high-redshift end than that in the re-sampling assuming only the
collisional distance constraint.

As shown in the right panels of
Figs.~\ref{f:zmbias}--\ref{f:zmbiaslimmag}
the scatter in the dynamical mass estimates for the clusters with at
least ten members is within 50\%, which is the minimum required
precision for a stacking analysis for the mass calibration given the
known intrinsic scatter in the mass--observable relation (e.g. Biviano
et al. 2006). We thus inspect the fraction of the clusters with at
least ten members. All of the 63 clusters in the SPIDERS and 4MOST
cases that assume no magnitude cut have more than ten galaxy members
per cluster. In the setups with the magnitude limit instead, the
fraction of the clusters with at least ten members per cluster
decreases with redshift as shown in Fig.~\ref{f:frac_mc}. In the
redshift bins of 0.2, 0.4, 0.6, and 0.8 in the SPIDERS\_55m setup,
there are one, three, eight, and 18 clusters out of 63 clusters, which
have fewer than ten redshifts per cluster. The case for the
SPIDERS\_65m setup, which uses a larger fiber separation
($65^{\prime\prime}$), is slightly worse. In the redshift bins of 0.4,
0.6 and 0.8 using the 4MOSTm setup, there are one, two and four
clusters out of 63 clusters, which have fewer than ten redshifts per
cluster.

\subsubsection{Redshift estimates}

We considered all clusters with $n_{\rm gal,MC}>1$ in the redshift
estimates. Here we do not consider $n_{\rm gal,MC}=1$ clusters
because the BCG is always taken according to the selection method
(Sect.~\ref{s:member_mc}) which leads to zero dispersion based on 500
MC realizations. The cluster redshifts are well recovered to
a level of a few per cent in all cases. There is no obvious systematic
dependence on the cluster redshift. The left panels of
Figs.~\ref{f:zmbias}--\ref{f:zmbiaslimmag} and
Fig.~\ref{f:zmbias_05zs} show the bias of the redshifts obtained in
the re-sampled procedure from the input cluster redshifts and its
dispersion based on the 500 MC realizations.  We note that the catalog
of the galaxy redshifts of the HIFLUGCS is a rather clean member
galaxy input catalog, likely almost free of interlopers. Uncertainties
derived from the scatter in the down-sampling thus do not account for
the effect of interlopers. This artificially restricts the
uncertainties of the redshift estimates to less than 0.01. In
Figs.~\ref{f:zmbias_10zs} and \ref{f:zmbias_05zs}, the scatter increases
slightly with the number of the input galaxies, which may indicate that
the scatter of redshift with a fixed number of spectroscopic members
depends on the richness.  Richness-dependent values of minimum numbers
of galaxy redshifts for different richness populations may thus be
required for a stacking analysis in order to ensure an equivalent
scatter introduced by individual systems in the redshift estimates.

\subsubsection{Dynamical mass estimates}

\begin{figure*}
\begin{center}
\vspace{-0.8cm}

\hspace{-1.5cm}
\includegraphics[angle=90,width=9cm]
{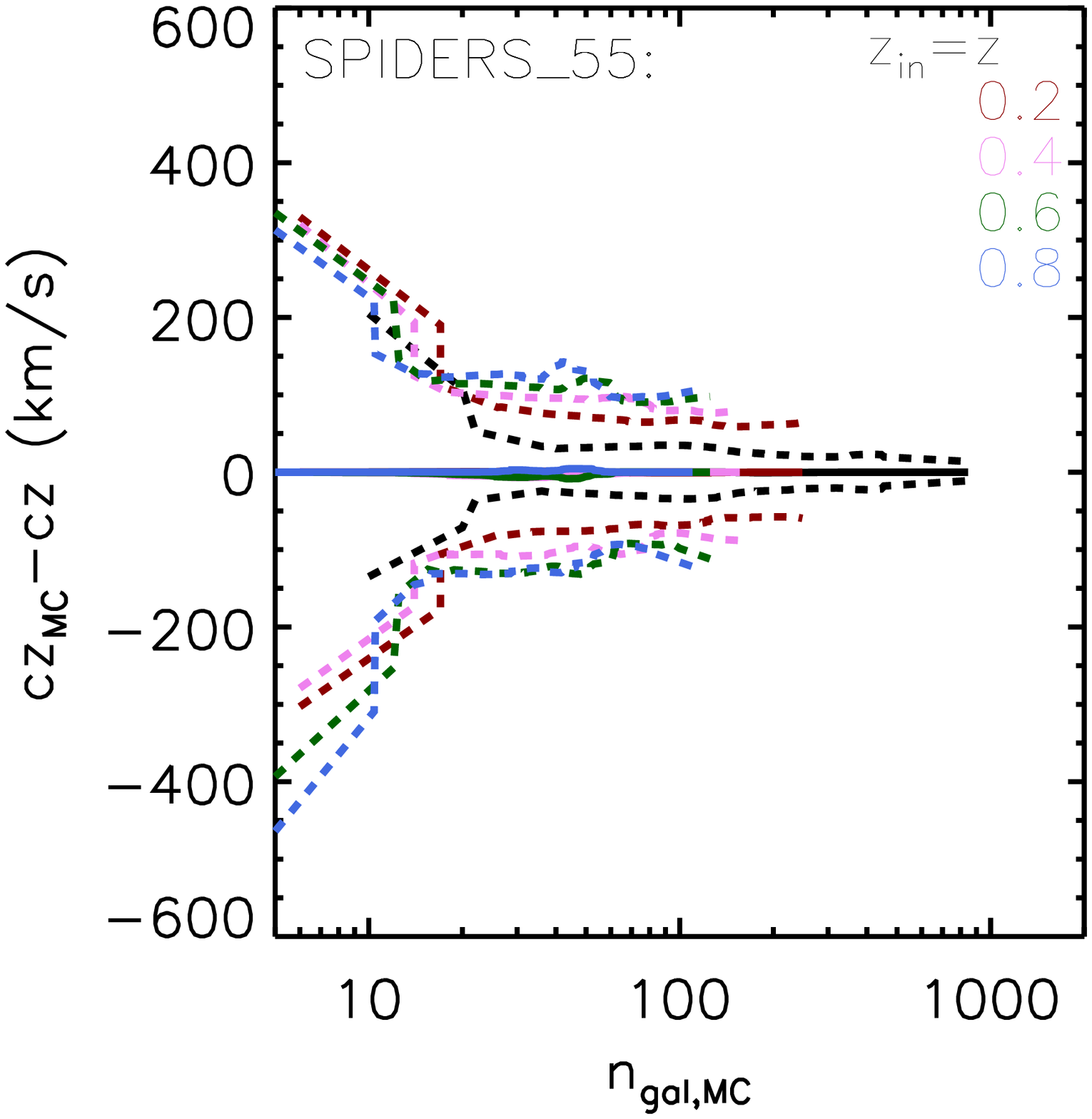}
\hspace{-2cm}
\includegraphics[angle=90,width=9cm]
{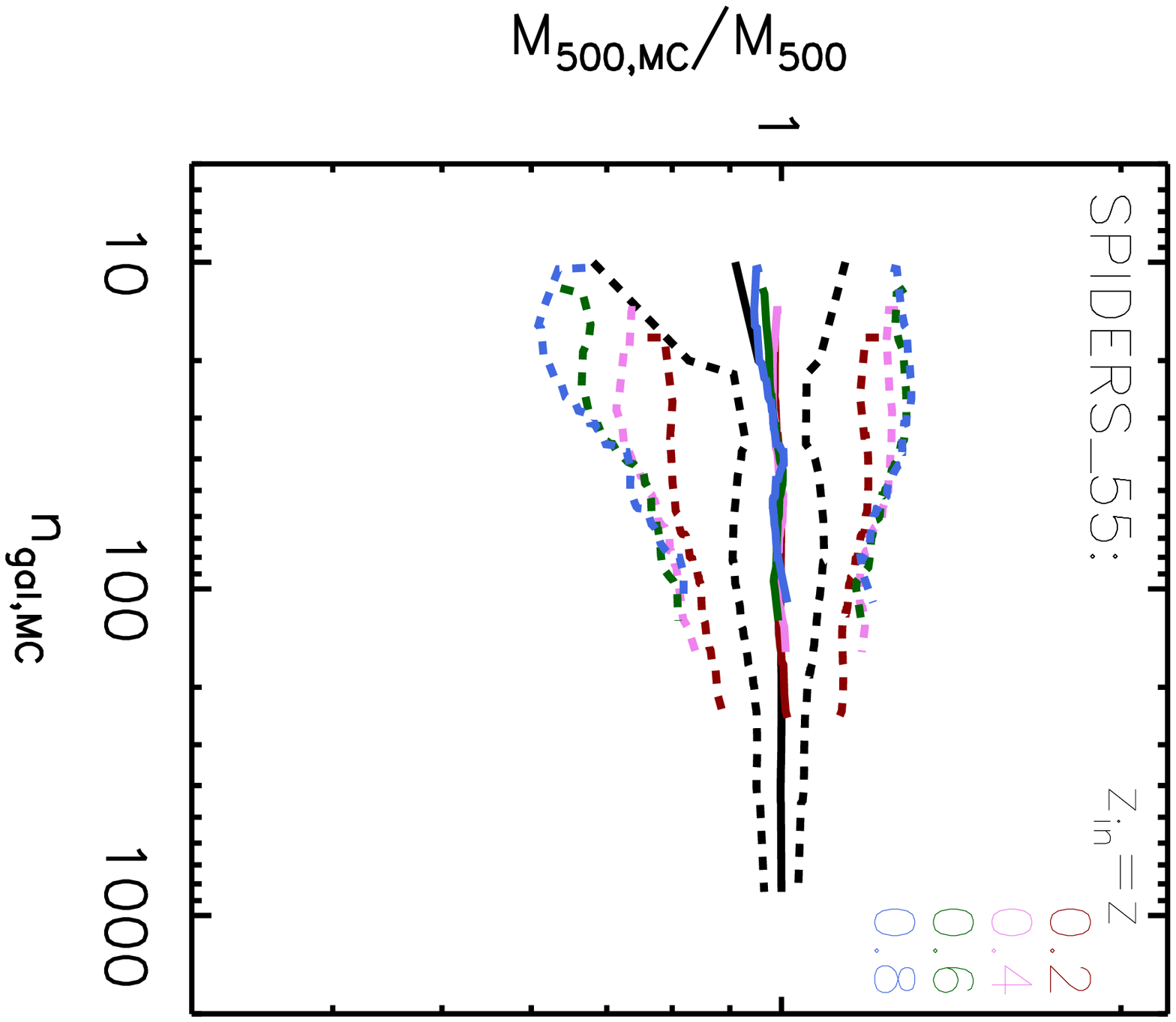}
\hspace{-1.5cm}
\vspace{-0.8cm}

\hspace{-1.5cm}
\includegraphics[angle=90,width=9cm]
{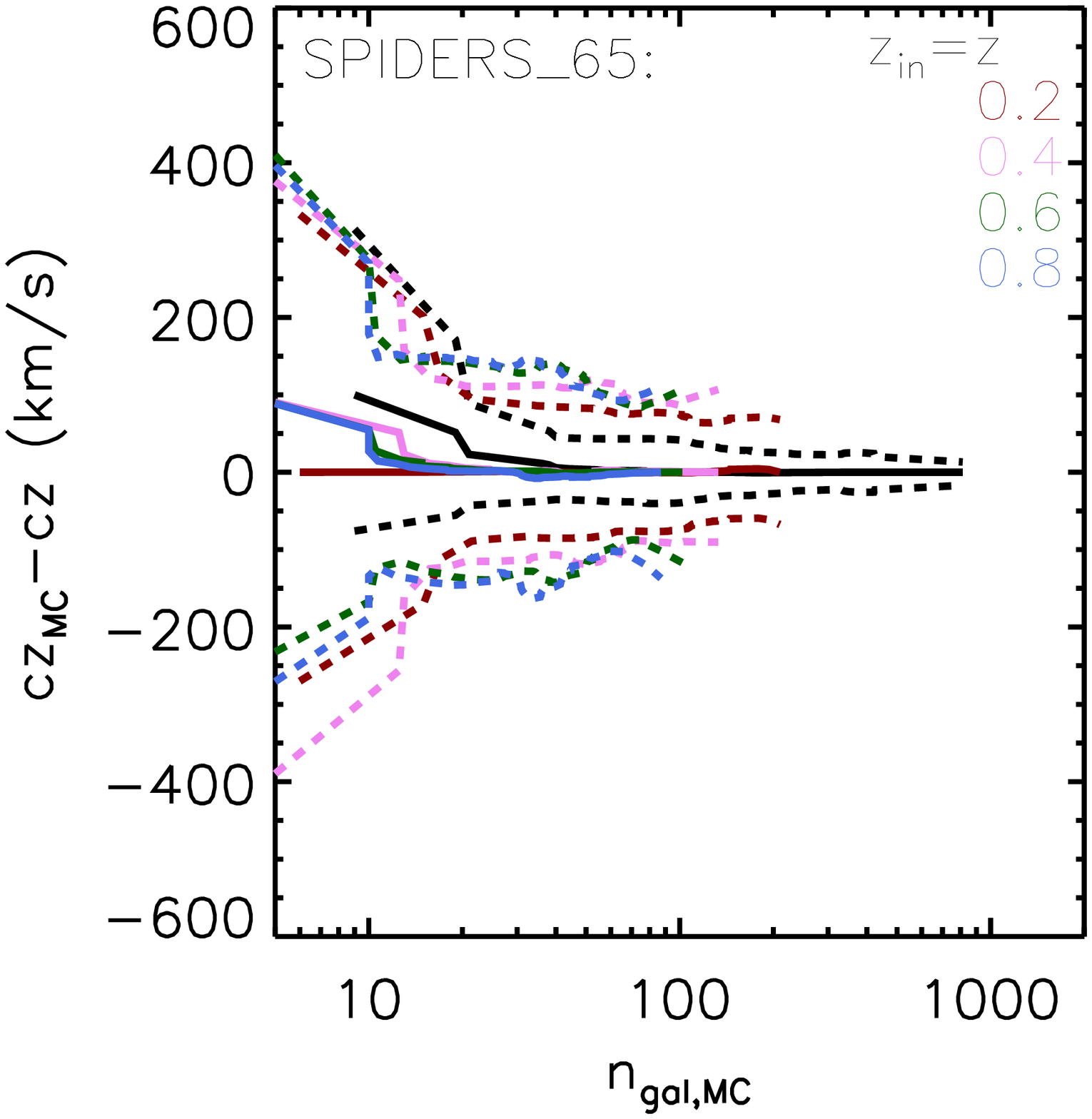}
\hspace{-2cm}
\includegraphics[angle=90,width=9cm]
{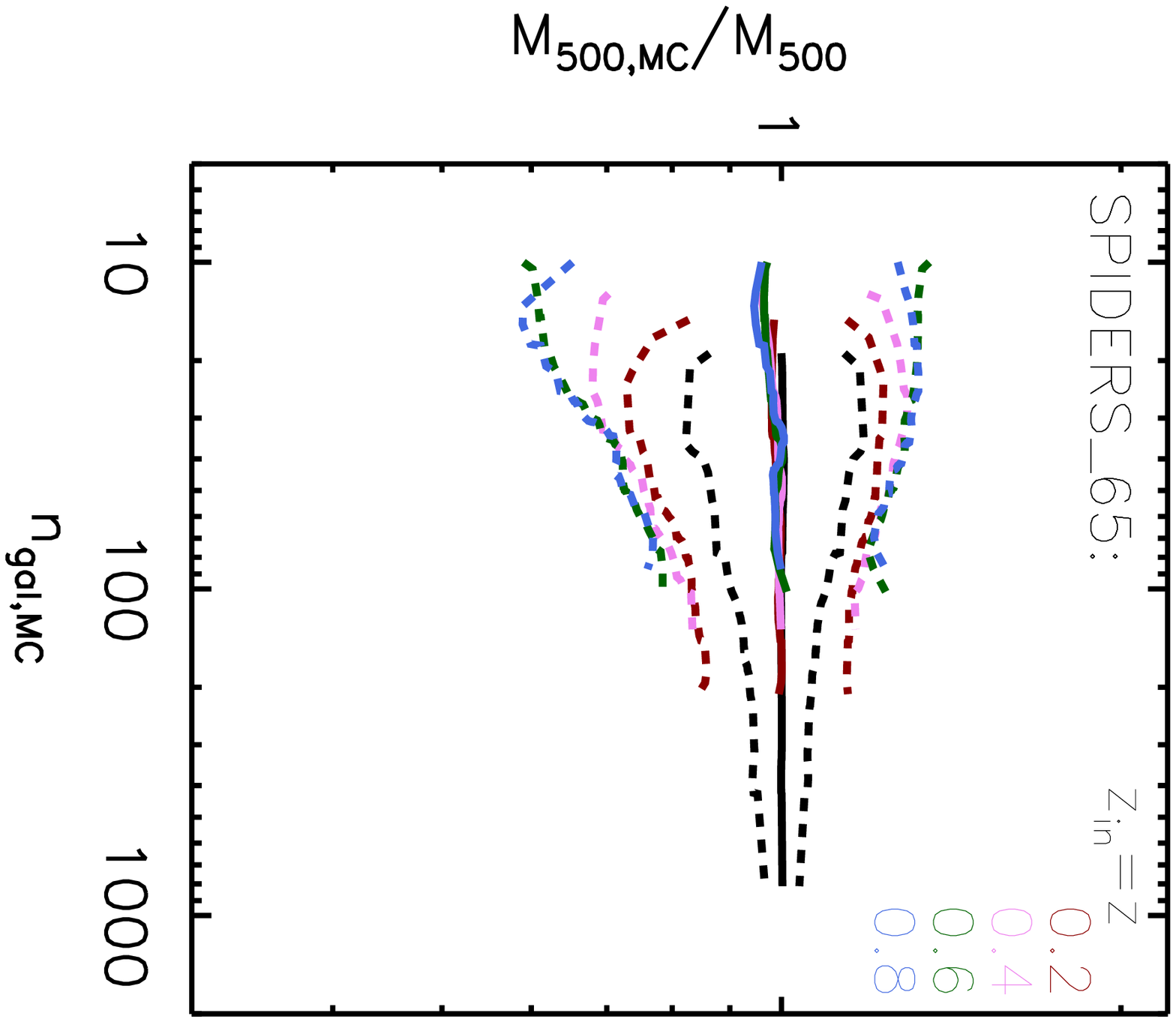}
\hspace{-1.5cm}
\vspace{-0.8cm}

\hspace{-1.5cm}
\includegraphics[angle=90,width=9cm]
{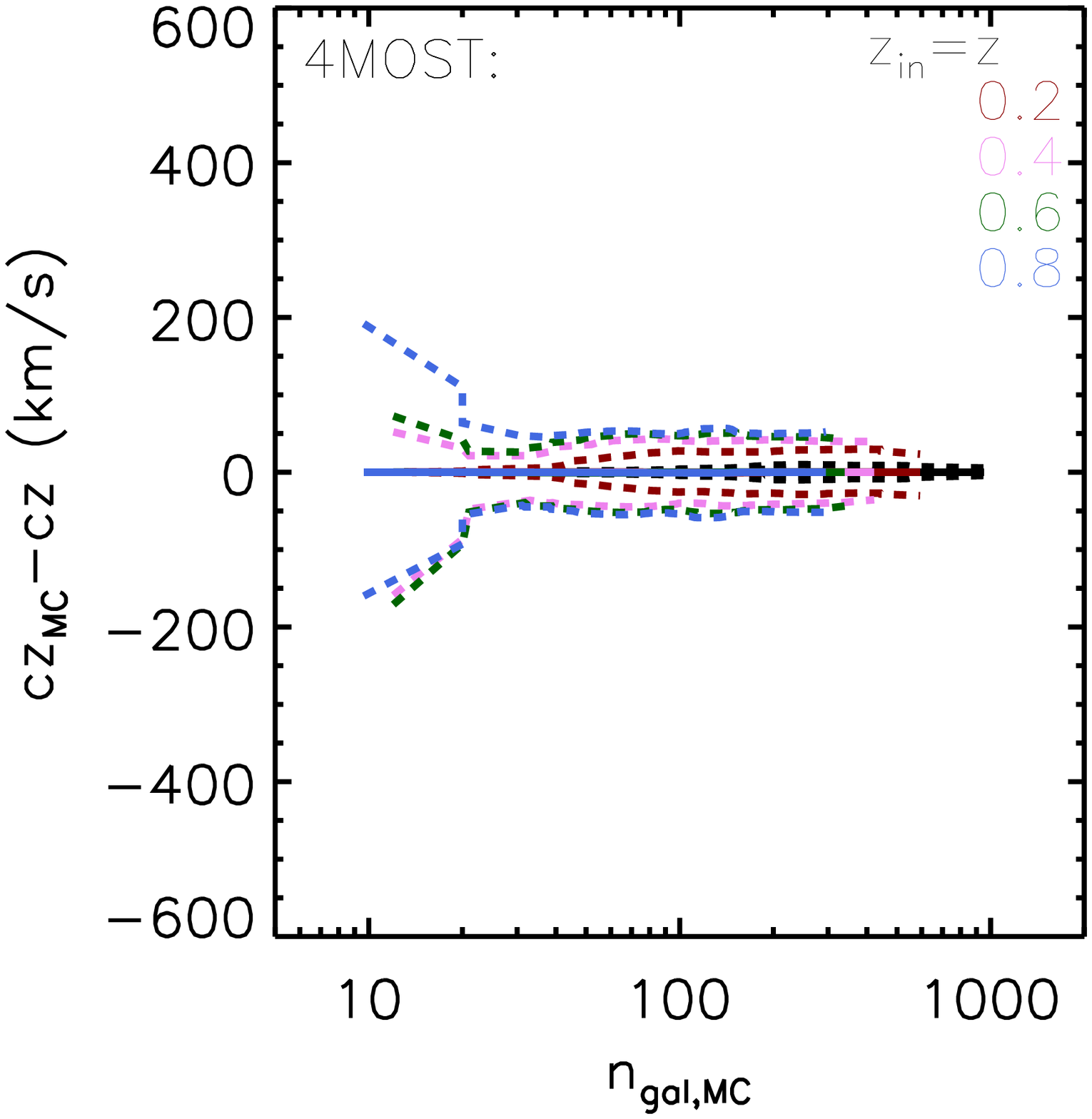}
\hspace{-2cm}
\includegraphics[angle=90,width=9cm]
{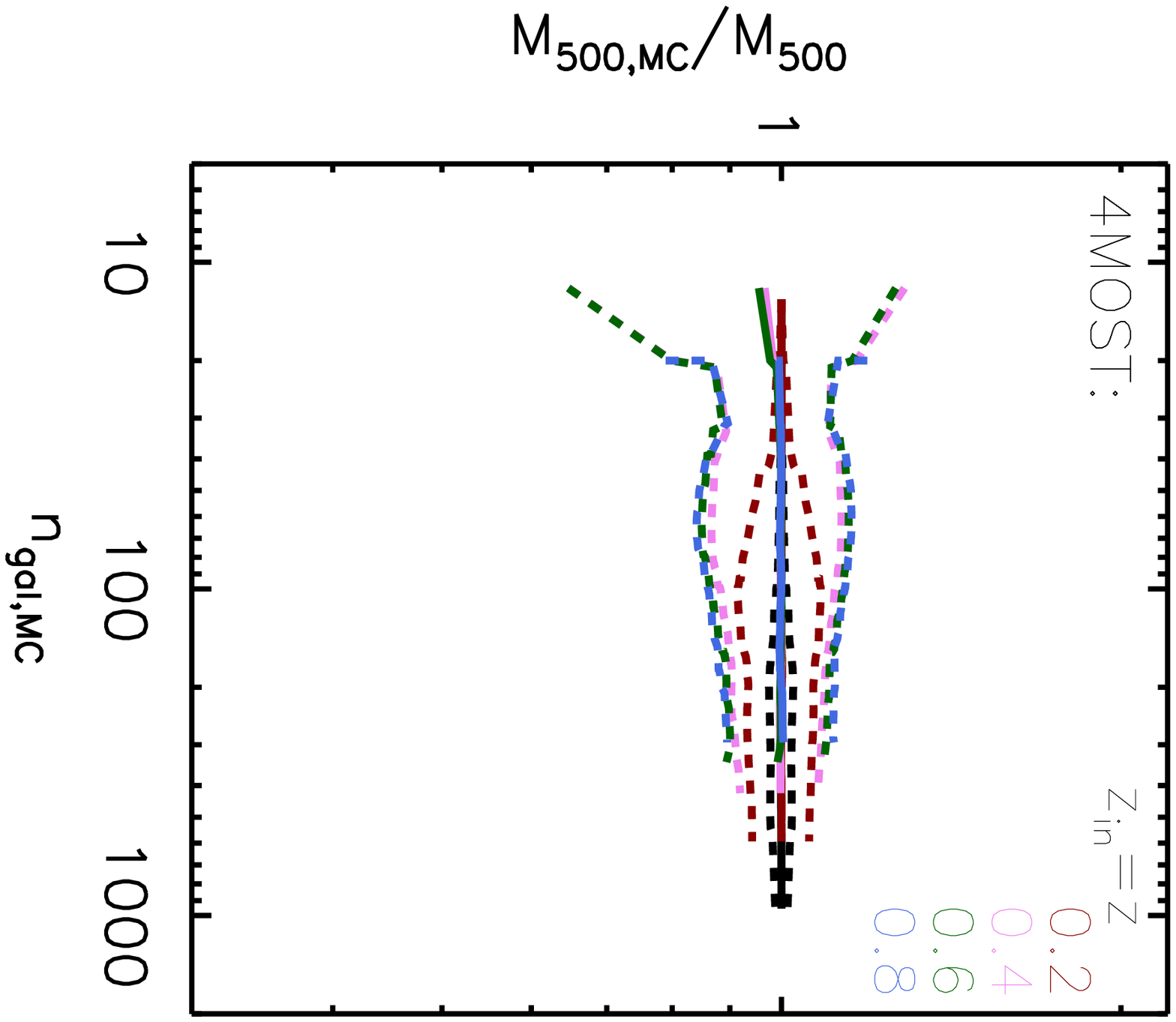}
\hspace{-1.5cm}
\vspace{-0.8cm}
\end{center}
\caption{Bias (solid lines) and dispersion (dashed lines) of the
  redshift (left panels) and dynamical mass (right panels)
  measurements in the SPIDERS\_55 (upper panels), SPIDERS\_65 (middle
  panels) and 4MOST (lower panels) setups from the input cluster
  values as a function of the re-sampled number of redshifts in
  use. The curves are smoothed with a boxcar average of the specified
  width of 11 to avoid spikes due to under-sampling for a few clusters.}
\label{f:zmbias}
\end{figure*}

\begin{figure*}
\begin{center}
\hspace{-1.5cm}
\includegraphics[angle=90,width=9cm]
{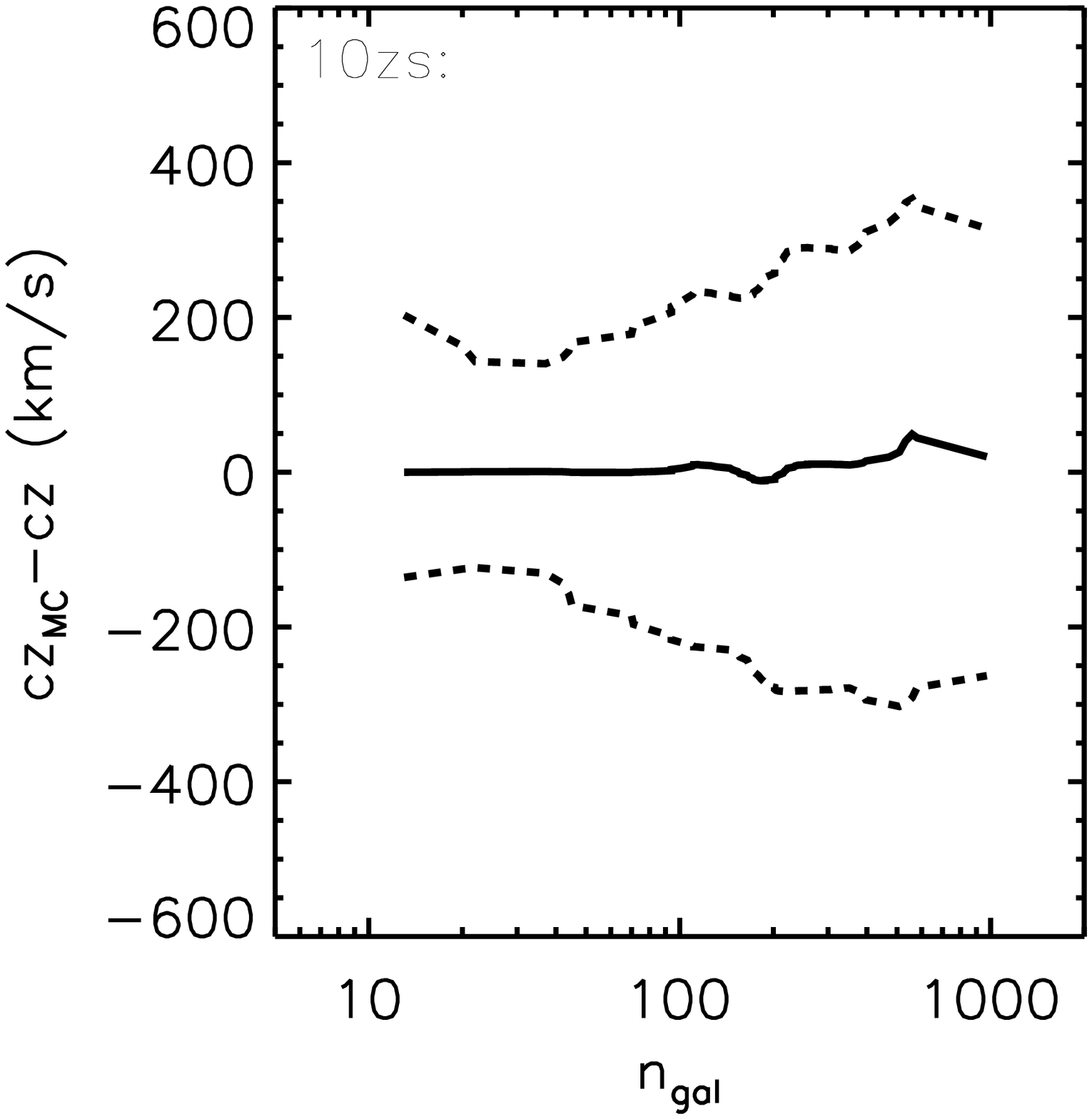}
\hspace{-2cm}
\includegraphics[angle=90,width=9cm]
{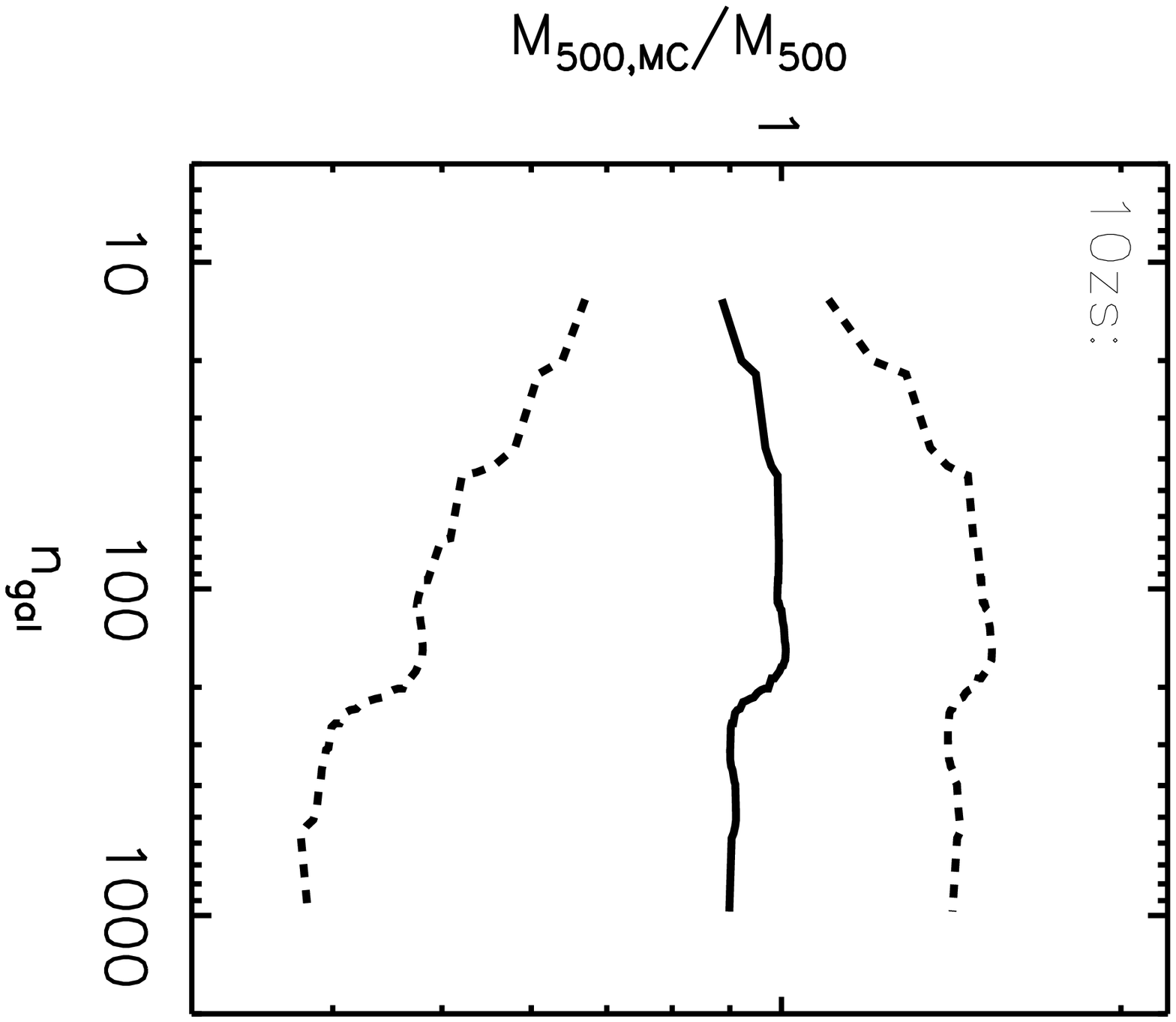}
\hspace{-1.5cm}
\end{center}
\caption{Bias (solid curves) and dispersion (dashed curves) of the
  redshift (left panel) and dynamical mass (right panel) measurements
  in the 10zs setup from the input cluster values versus input number
  of cluster galaxies. The curves are smoothed with a boxcar average
  of the specified width of 11 to avoid
  spikes due to under-sampling for a few clusters. Note that $n_{\rm
    gal}$ is the input number of galaxies and the output number of
  galaxies is always 10.  }
\label{f:zmbias_10zs}
\end{figure*}

\begin{figure*}
\begin{center}
\vspace{-0.8cm}
\hspace{-1.5cm}
\includegraphics[angle=90,width=9cm]
{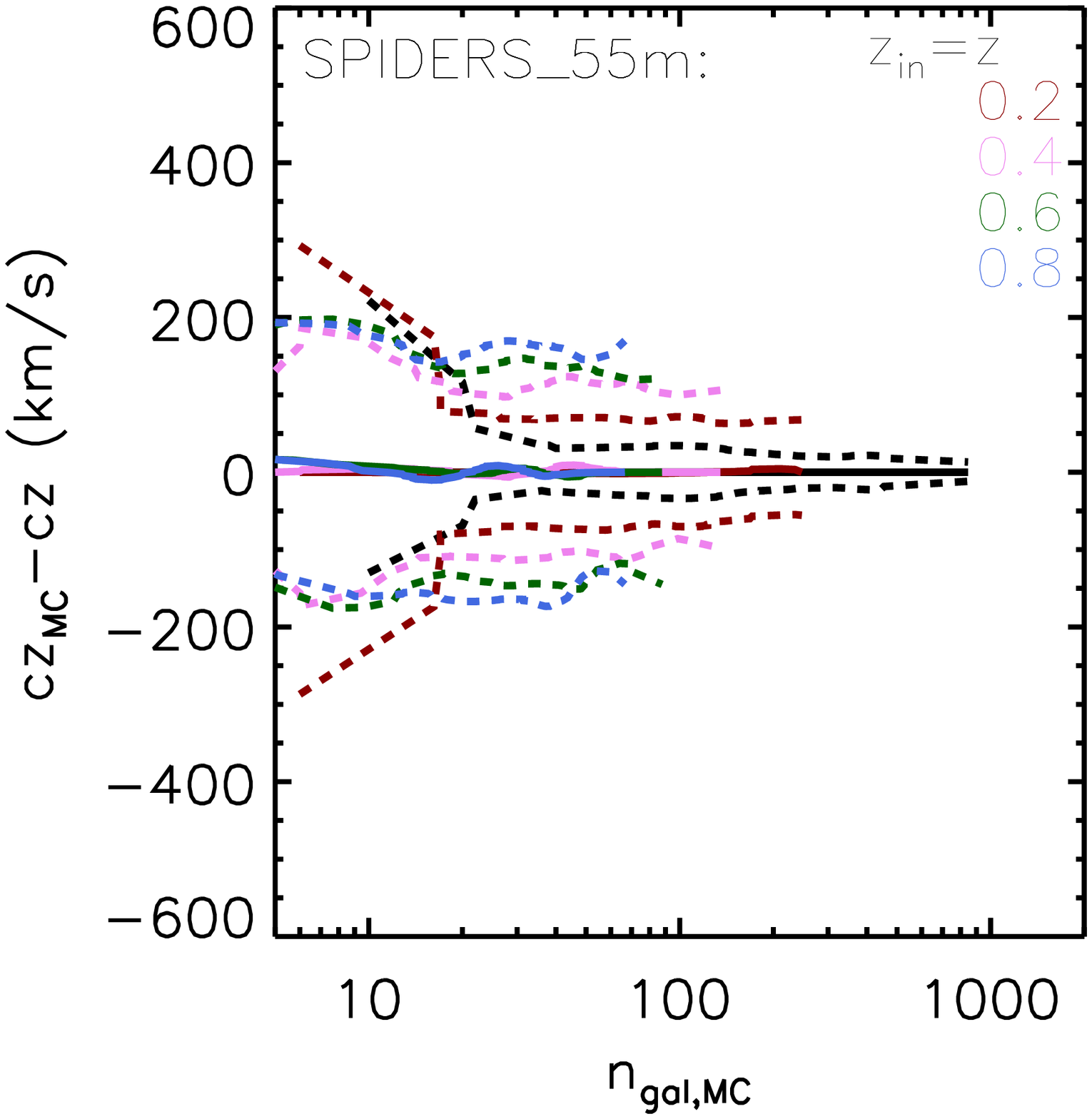}
\hspace{-2cm}
\includegraphics[angle=90,width=9cm]
{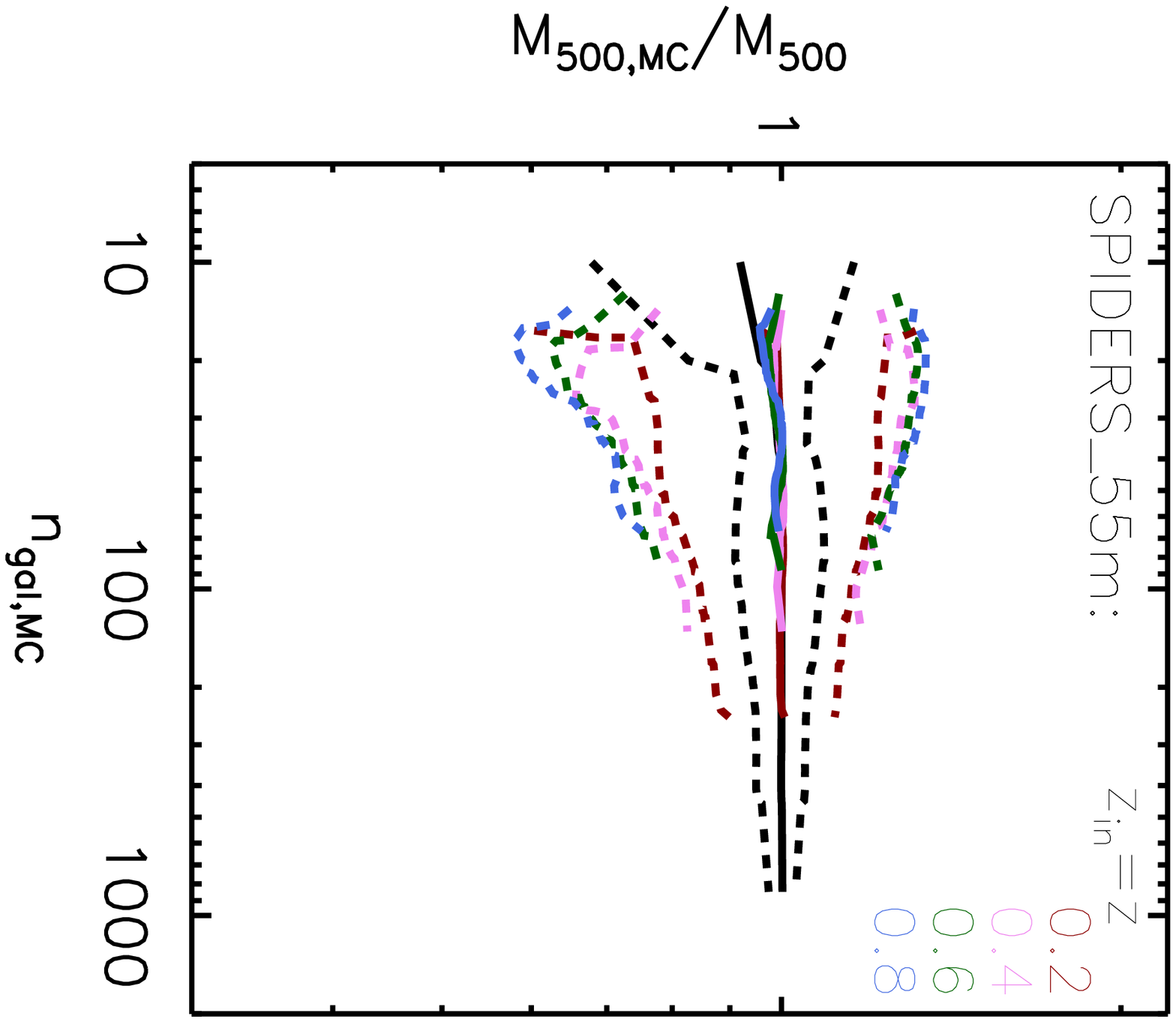}
\hspace{-1.5cm}

\vspace{-0.8cm}
\hspace{-1.5cm}
\includegraphics[angle=90,width=9cm]
{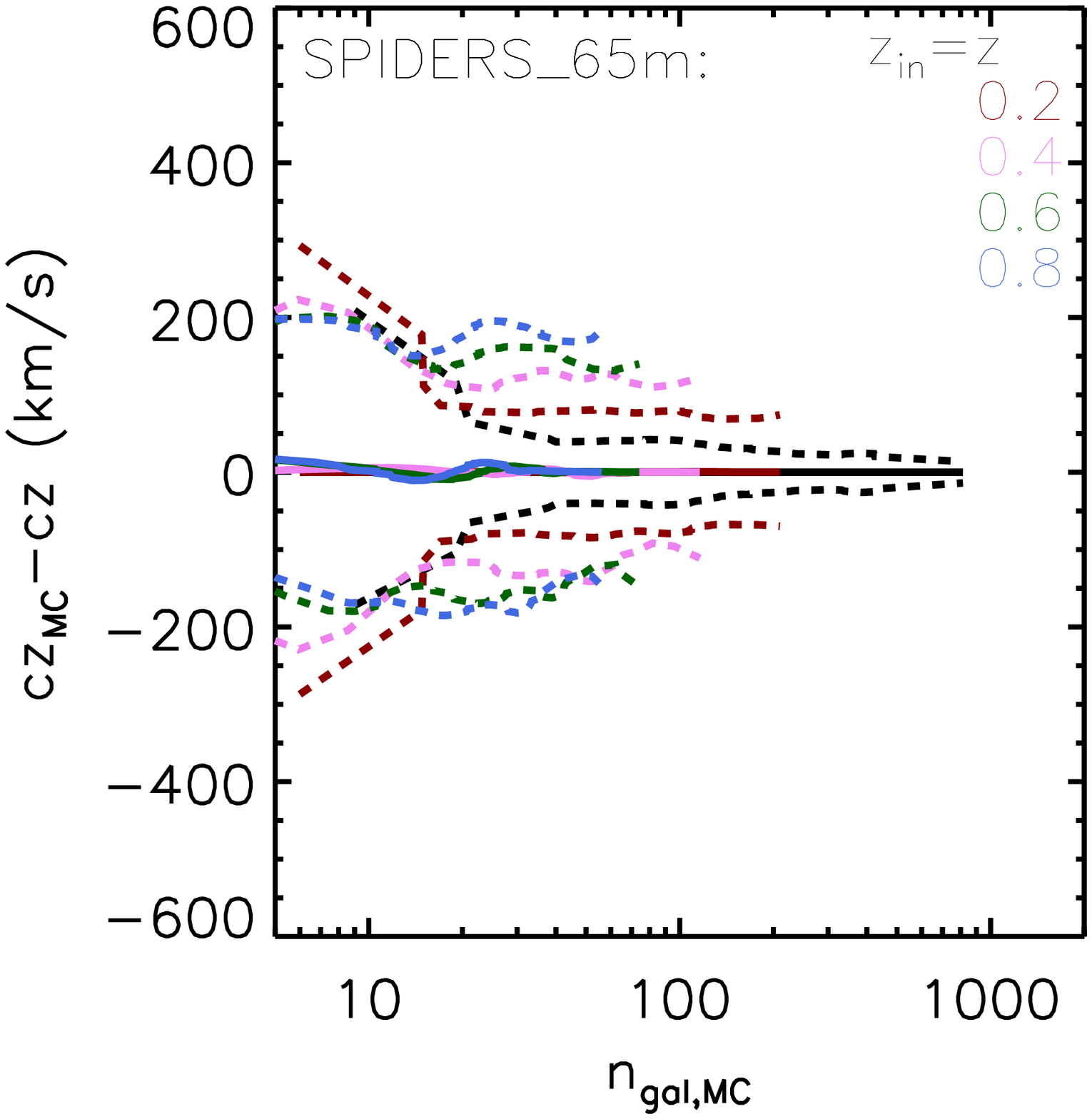}
\hspace{-2cm}
\includegraphics[angle=90,width=9cm]
{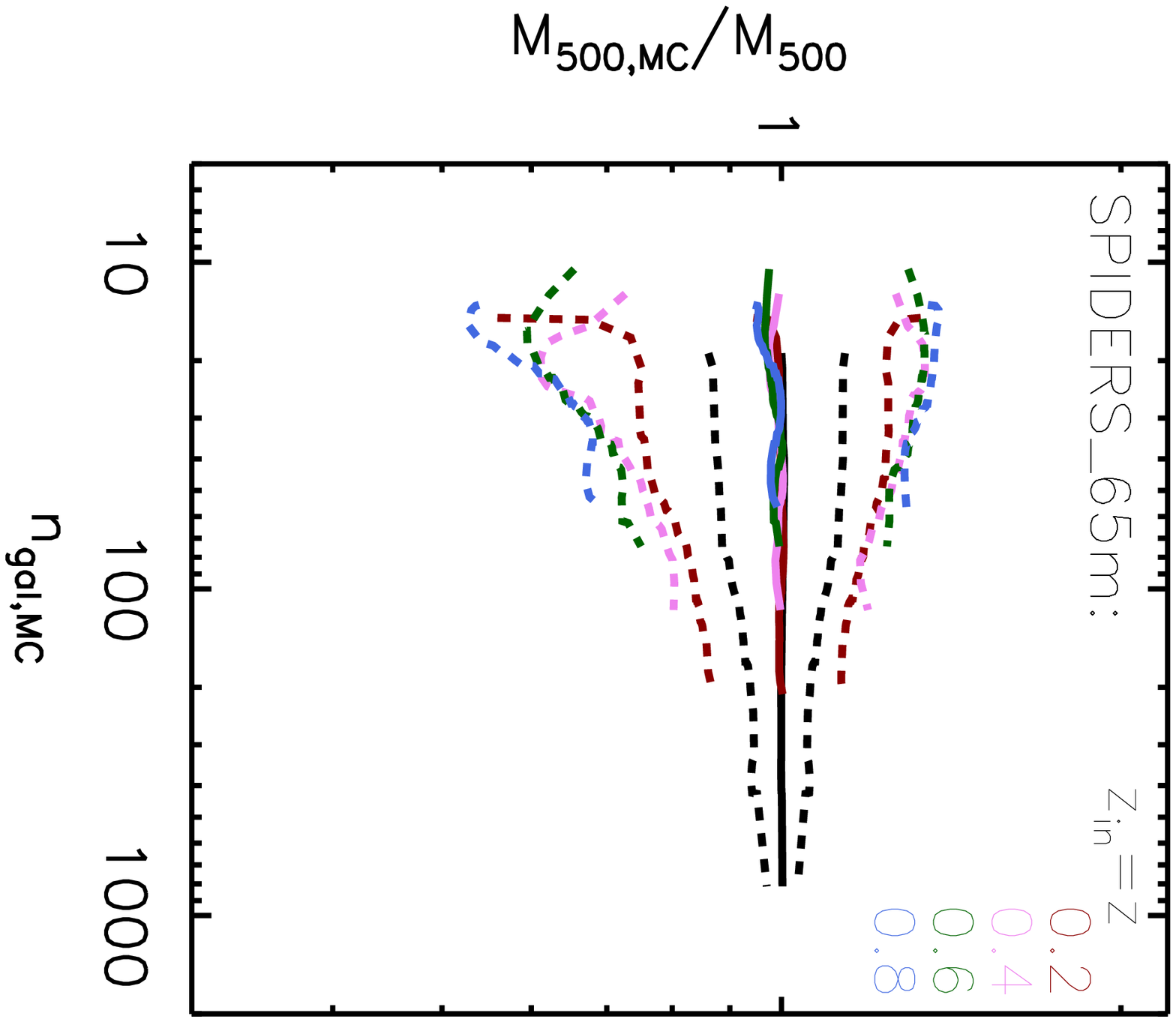}
\hspace{-1.5cm}

\vspace{-0.8cm}
\hspace{-1.5cm}
\includegraphics[angle=90,width=9cm]
{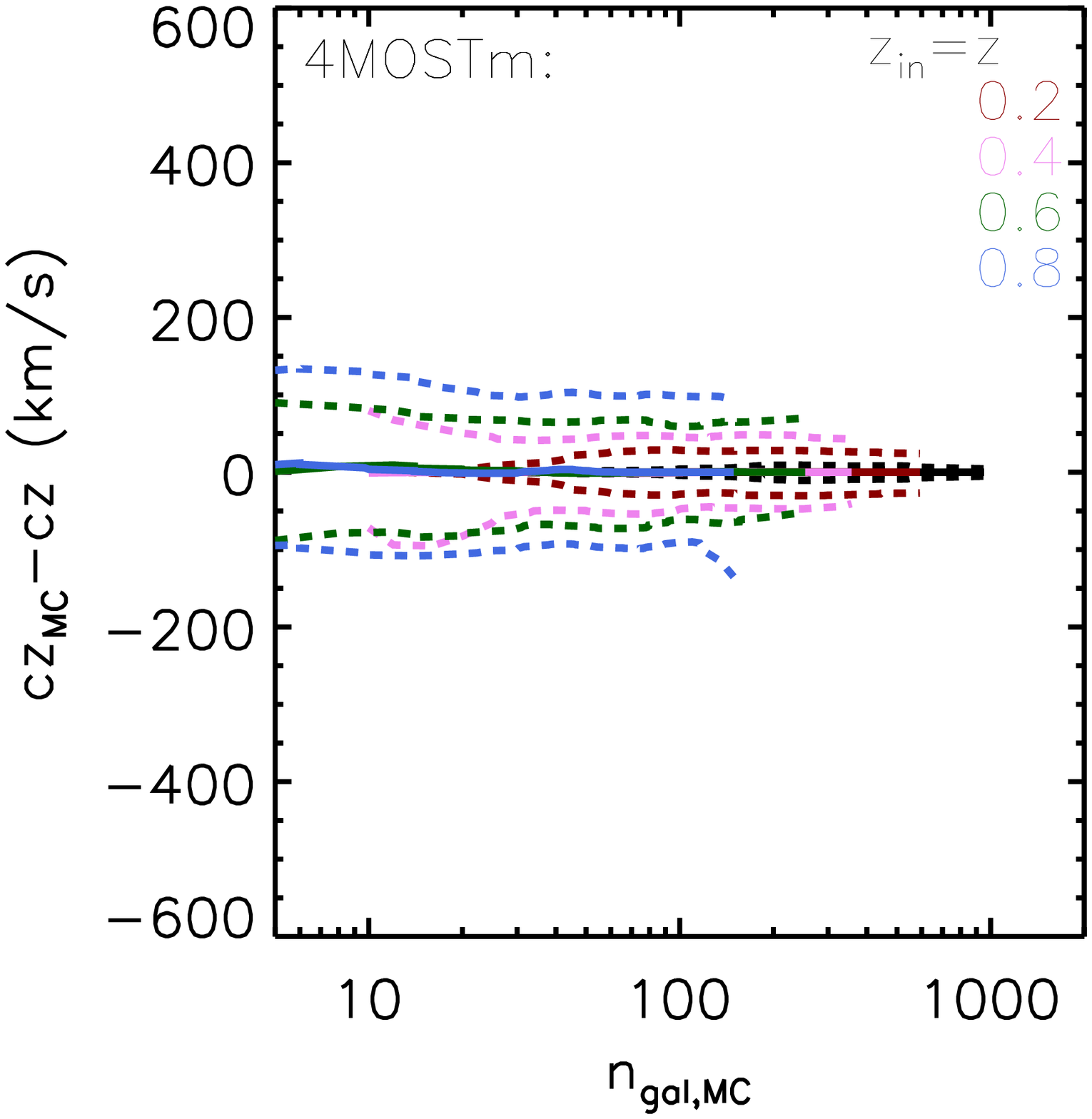}
\hspace{-2cm}
\includegraphics[angle=90,width=9cm]
{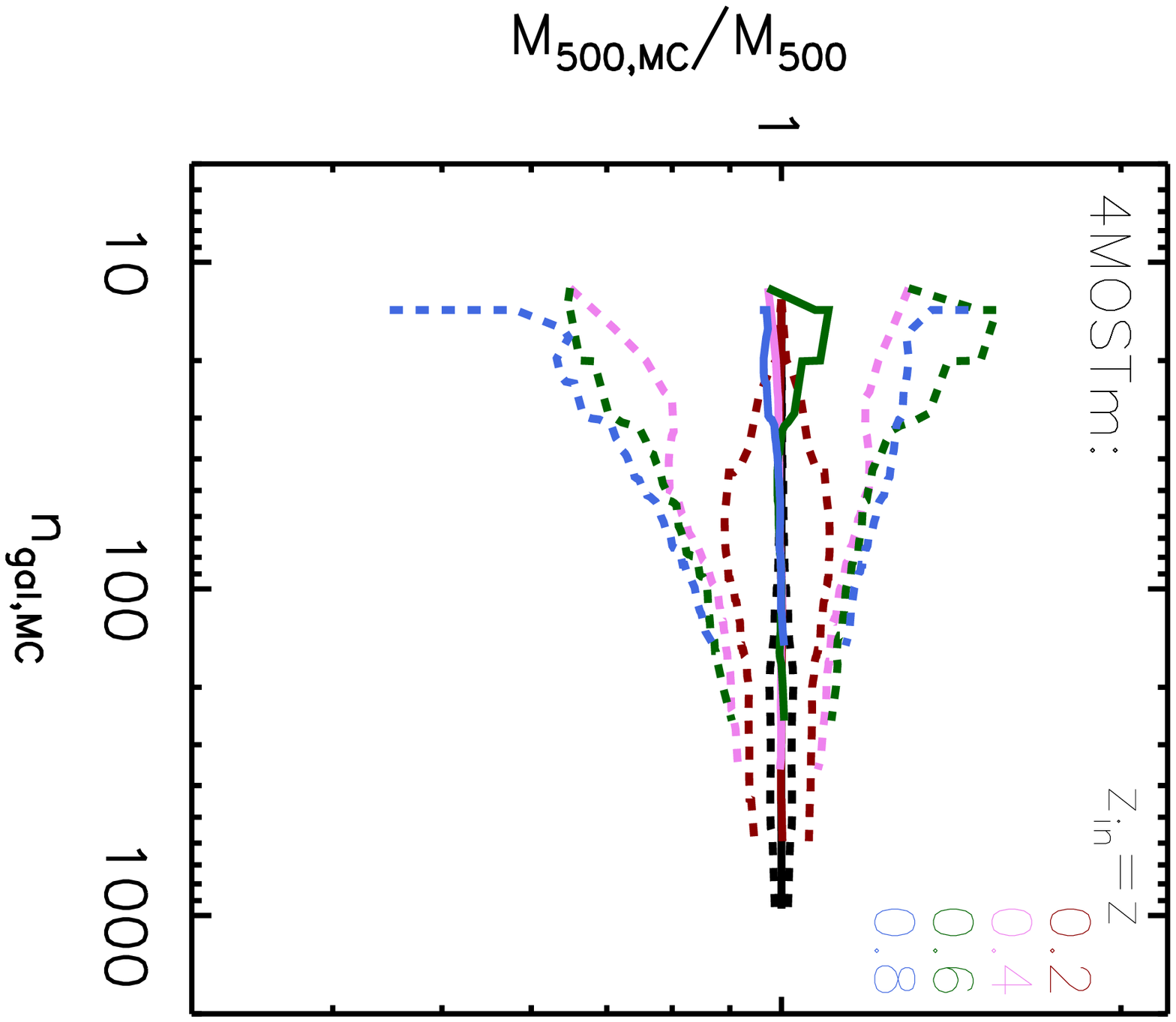}
\hspace{-1.5cm}
\vspace{-0.8cm}
\end{center}
\caption{Bias (solid curves) and dispersion (dashed curves) of the
  redshift (left panels) and dynamical mass (right panels)
  measurements in the SPIDERS\_55m (upper panels),
  SPIDERS\_65m (middle panels) and 4MOSTm (lower panels) setups
  from the input cluster values as a function of the
  re-sampled number of redshifts in use. The curves are smoothed with 
  a boxcar average of the
  specified width of 11 to avoid spikes due
  to under-sampling for a few clusters.
}
\label{f:zmbiaslimmag}
\end{figure*}

\begin{figure*}
\begin{center}
\includegraphics[angle=90,width=15cm]{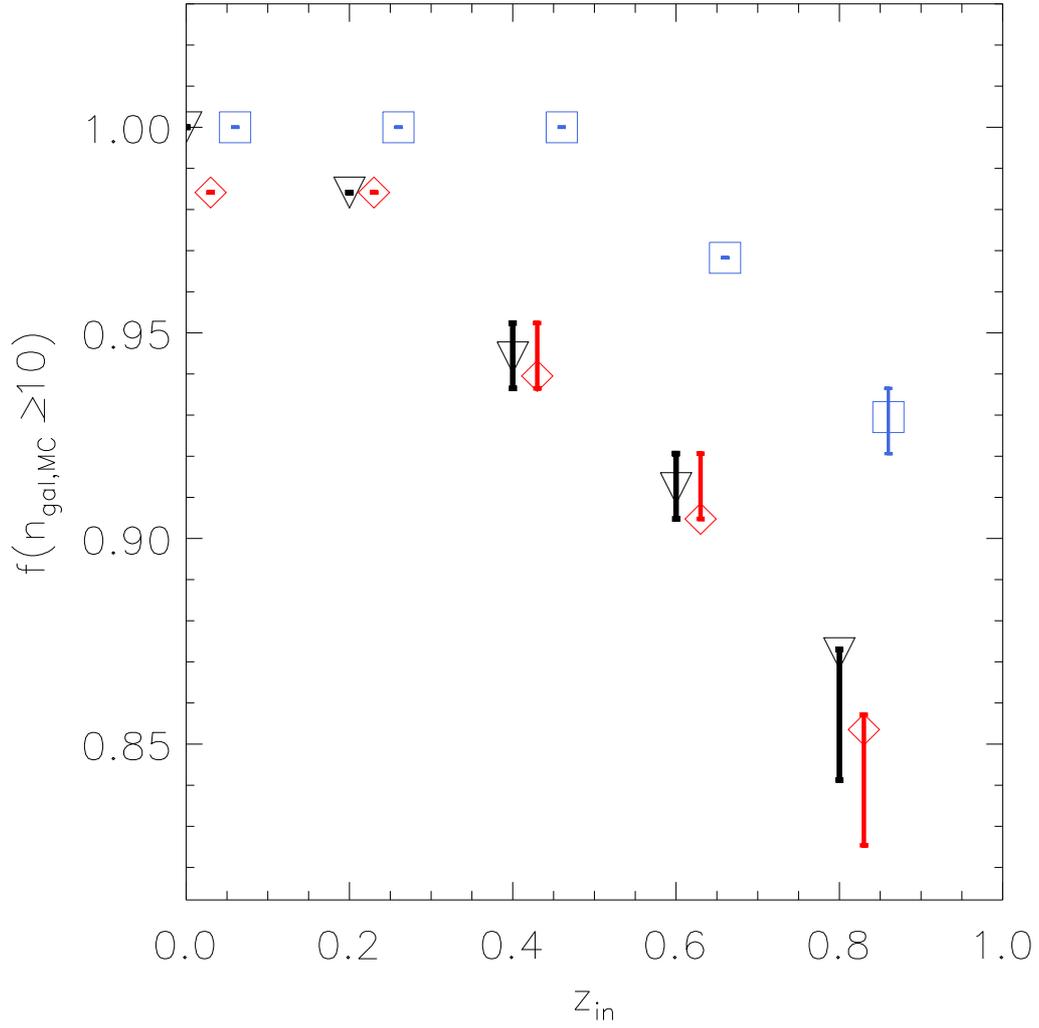}
\end{center}
\caption{Fraction of the clusters with at least ten redshifts per
  cluster after the re-sampling using the SPIDERS\_55m (black
  triangles), SPIDERS\_65m (red diamonds, with 0.03 offset) and
  4MOSTm (blue squares, with 0.06 offset) setups as a function of
  the assigned input cluster redshift ($z_{\rm in}$) in the
  re-sampling. Note that $z_{\rm in}=0$ case refers to the re-sampled
  clusters at the cluster redshifts $z$.}
\label{f:frac_mc}
\end{figure*}

\begin{figure*}
\begin{center}
\includegraphics[angle=90,width=9cm]{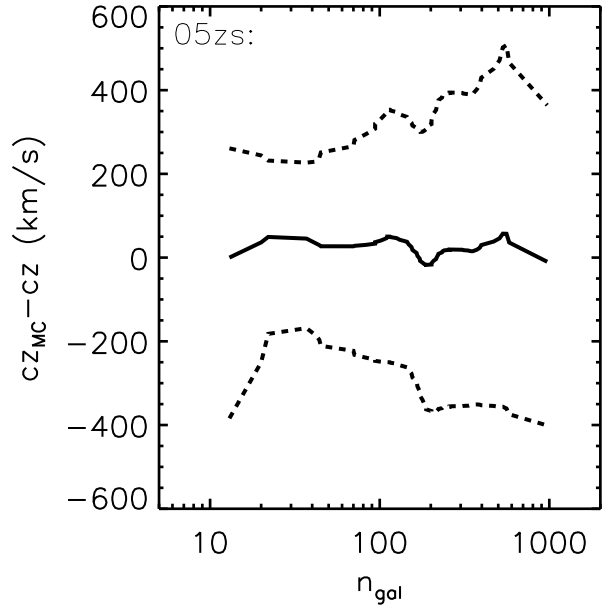}
\end{center}
\caption{Bias (solid curves) and dispersion (dashed curves) of the
  redshift measurements in the 05zs setup from the input cluster
  values versus input number of cluster galaxies. The curves are
  smoothed with a boxcar average of the specified width of 11 to avoid
  spikes due to under-sampling for a few clusters. Note that $n_{\rm
    gal}$ is the input number of galaxies and the output number of
  galaxies is always 5.  }
\label{f:zmbias_05zs}
\end{figure*}

There is no redshift or mass dependence in the systematic
uncertainties of the dynamical mass estimates. The right panels of
Figs.~\ref{f:zmbias}--\ref{f:zmbiaslimmag}
show the bias of the dynamical mass estimates obtained in the
re-sampled procedure from the cluster total mass and its dispersion
based on the 500 MC realizations. Here we do not consider $n_{{\rm
    gal},i}<10$ realizations as noted in
Sect.~\ref{s:zmestimator_mc}. Therefore, toward the low-$n_{\rm
  gal,MC}$ end, the bias and dispersion is slightly underestimated as
not all 500 MC realizations can be used.

The dynamical mass is well recovered with less than $\sim$20\% bias on
average. The dispersion in Figs.~\ref{f:zmbias}--\ref{f:zmbiaslimmag}
demonstrates the precision the corresponding setup can reach as a
function of redshift. This information can be used for the sample
selection corresponding to the required precision for the purpose of
mass calibration in upcoming surveys. Given the same required
accuracy, the 4MOST survey can provide mass calibration up to a higher
redshift compared to the SPIDERS survey. With ten redshifts per
cluster, the dynamical mass measurement can easily be biased by a
factor of two for individual clusters, which would not be suitable for
a robust mass calibration unless stacking them. Similar to that shown
in the redshift estimates, the scatter increases again slightly with
the number of the input galaxies, this may indicate a dependence on
richness. Richness dependent cuts of minimum numbers of galaxy
redshifts in different richness bins shall also be required for a
stacking analysis to ensure an equivalent scatter introduced
by individual systems in the dynamical mass estimates.

\subsubsection{Limitations}

The current paper describes an ideal situation of using an input
member catalog almost free from contamination from interlopers for the
spectroscopic follow-up that does not represent the reality in
general. According to real surveys, we need to construct our input
galaxy candidate sample for the spectroscopic follow-up, which can not
avoid contamination from, for example, interlopers. Additionally, we
need to select a number of certain member galaxies from a large
catalog for the follow-up to determine the redshift and velocity
dispersion, which requires knowledge of the selection effects for
recovering the mass estimates robustly. The input member galaxy sample
for the simulation in the current paper, however, likely contains no
interlopers. We also can not quantify the selection effect as the
sample is a collection of clusters from the available literature,
instead of e.g. a spectroscopic follow-up of a well-defined
photometric galaxy member sample. A further study based on mock data
in simulations with known selection functions as well as sufficient
fore- and background galaxies in the galaxy sample will help to
address these questions. Another limitation is that clusters may
evolve from $z=0.8$ to the present epoch, an effect that is not
  fully accounted for by simply shifting the HIFLUGCS clusters to
higher redshifts. It is thus not obvious that the HIFLUGCS sample is
the best test-case for the forecasts of the 4MOST and SPIDERS surveys
at higher redshifts.

%

\section{Conclusions} 
\label{s:con}

We analyzed $\sim 1.3$~Ms of clean \XMM data and \ROSAT pointed
observations as well as optical spectroscopic redshifts of 13650
cluster member galaxies for all 64 HIFLUGCS clusters. Excluding
2A~0335+096 with only three redshifts, we present the $L^{\rm cor}-M$
relation for 63 nearby clusters of galaxies in the HIFLUGCS. For the
optimal use of the optical spectroscopic surveys for high-redshift
galaxy clusters and groups observed in upcoming X-ray surveys for mass
calibrations, we carried out MC re-sampling of the galaxy members with
spectroscopic redshifts, and calibrated the systematic uncertainties
in the redshift and dynamical mass estimates. We predicted the
redshift and dynamical mass estimates assuming the SPIDERS and 4MOST
optical spectroscopic survey setups, respectively, using MC
re-sampling of the 63 HIFLUGCS clusters by placing the sample at the
actual cluster redshifts as well as at the redshifts of 0.2, 0.4, 0.6,
and 0.8. Aiming for high-redshift cluster/group systems, we also
predicted the redshift estimates based on five and ten spectroscopic
members per cluster, respectively, and the mass estimates based on ten
spectroscopic members per cluster. We list the conclusions in detail
as follows.

\begin{itemize}


\item Given sufficient numbers (i.e. $\ge$45) of member galaxies in
  computing the dynamical masses, the $L^{\rm cor}-M$ relations agree
  between the disturbed and undisturbed clusters.

\item The CC clusters still dominate the scatter in the $L^{\rm cor}-M$
  relation even when the CC corrected X-ray luminosity is
  used. This indicates that the scatter of the $L^{\rm cor}-M$ scaling relation
  mainly reflects the structure formation history of the clusters.


\item The dynamical masses can be measured within 10\% uncertainty
  using the SPIDERS\_55, SPIDERS\_65 and 4MOST setups which is
  independent of cluster redshift and mass. With ten redshifts per
  cluster or more, the dynamical masses can be
  recovered with less than 20\% bias on average, in which the
  dynamical masses are underestimated for most systems.

\item The bias of the cluster dynamical mass estimates increases
  toward the high-redshift end. The underestimation of the cluster
  masses on average using the SPIDERS\_65m setup, in which both
  the collisional distance of 65$^{\prime\prime}$ and magnitude cut are
  considered, is better than 19\%,
  28\%,
  34\%, 
  and 37\% 
  in the redshift bins of 0.2, 0.4, 0.6, and 0.8. The dynamical mass
  is recovered with less than 20\% underestimation up to redshift 0.6
  using the 4MOSTm setup, in which both the collisional distance and
  magnitude cut are considered. In the redshift bin of 0.8, the
  underestimation of the dynamical masses on average, according to the
  4MOSTm setup, is still less than 24\%.
  Assuming the SPIDERS\_55m (4MOSTm) setup, the dynamical masses can
  be used as an independent reference blind to the X-ray observables
  to calibrate the cluster mass with less than 20\% underestimation up
  to redshift 0.2 (0.6) with 2\% (3\%) catastrophic outliers
  (i.e. fewer than ten members per cluster) in upcoming X-ray
  surveys. Empirically, one can correct this bias using a complete
  sample from observations or a mock sample in simulations, which
  includes the causes of the bias such as contamination from fore- and
  background galaxies.

\end{itemize}

\begin{acknowledgements}

  The \XMM project is an ESA Science Mission with instruments and
  contributions directly funded by ESA Member States and the USA
  (NASA). The \XMM project is supported by the Bundesministerium f\"ur
  Wirtschaft und Technologie/Deutsches Zentrum f\"ur Luft- und
  Raumfahrt (BMWi/DLR, FKZ 50 OX 0001) and the Max-Planck
  Society. This research has made use of the NASA/IPAC Extragalactic
  Database (NED) which is operated by the Jet Propulsion Laboratory,
  California Institute of Technology, under contract with the National
  Aeronautics and Space Administration. We thank Tom Dwelly for
  helpful discussion and an anonymous referee for his/her insight
    and expertise that improved the work. Y.Y.Z. acknowledges support
  by the German BMWi through the Verbundforschung under grant
  50\,OR\,1506. T.H.R. acknowledges support from the DFG through the
  Heisenberg research grant RE 1462/5 and grant RE 1462/6.

\end{acknowledgements}

\bibliography{Abell}


\clearpage

\begin{table*}
\begin{center}
\renewcommand{\arraystretch}{0.92}
\caption[]{Number of spectroscopic redshifts, velocity dispersion,
  dynamical mass, core-corrected X-ray bolometric luminosity,
  luminosity concentration and offset between the BCG and X-ray
  flux-weighted centroid for the whole HIFLUGCS sample of 64 clusters
  sorted by R.A. as shown in Table~1 in Zhang et al. (2011).}
  \begin{tabular}{l|rrrrrrr}
\hline
\hline
    Name & $n_{\rm gal}$ & $\sigma$ (km/s) & $M_{500}$ ($10^{14}M_{\odot}$)          & $L^{\rm cor}_{500} $ ($10^{44}$~erg/s)  & $c_{\rm L}$  &  ${\Delta}R/r_{500}$ \\
    \hline
    A0085              &    350   & $  963  \pm    39 $ & $     4.83  \pm      0.76 $ & $     7.60   \pm     0.43 $ &      2.089   &      0.002 \\  
    A0119              &    339   & $  797  \pm    38 $ & $     2.67  \pm      0.47 $ & $     3.03   \pm     0.21 $ &      5.383   &      0.014 \\  
    A0133              &    137   & $  725  \pm    44 $ & $     1.98  \pm      0.42 $ & $     1.45   \pm     0.09 $ &      1.809   &      0.027 \\  
    NGC0507            &    110   & $  503  \pm    33 $ & $     0.68  \pm      0.15 $ & $     0.11   \pm     0.01 $ &      2.062   &      0.014 \\  
    A0262              &    138   & $  527  \pm    30 $ & $     0.78  \pm      0.16 $ & $     0.48   \pm     0.06 $ &      2.705   &      0.006 \\  
    A0400              &    114   & $  647  \pm    40 $ & $     1.44  \pm      0.30 $ & $     0.39   \pm     0.03 $ &      3.499   &      0.005 \\  
    A0399              &    101   & $ 1223  \pm    75 $ & $    10.37  \pm      2.18 $ & $     5.63   \pm     0.48 $ &      2.415   &      0.056 \\  
    A0401              &    116   & $ 1144  \pm    74 $ & $     8.32  \pm      1.82 $ & $    11.76   \pm     0.91 $ &      2.248   &      0.015 \\  
    A3112              &    111   & $  740  \pm    63 $ & $     2.09  \pm      0.57 $ & $     3.68   \pm     0.19 $ &      1.569   &      0.024 \\  
    NGC1339/Fornax     &    339   & $  366  \pm    13 $ & $     0.27  \pm      0.04 $ & $     0.03   \pm     0.00 $ &      2.595   &      0.001 \\  
 2A~0335$+$096         &     3    &   ---               &   ---                       & $     1.43   \pm     0.12 $ &      0.130   &      0.006 \\
    III~Zw~054         &     45   & $  657  \pm    62 $ & $     1.49  \pm      0.45 $ & $     0.43   \pm     0.06 $ &      2.042   &      0.025 \\  
    A3158              &    258   & $ 1044  \pm    45 $ & $     6.25  \pm      1.03 $ & $     5.41   \pm     0.41 $ &      2.453   &      0.032 \\  
    A0478              &     13   & $  945  \pm   223 $ & $     4.45  \pm      3.20 $ & $    12.26   \pm     0.69 $ &      1.572   &      0.003 \\  
    NGC1550            &     22   & $  263  \pm    34 $ & $     0.11  \pm      0.04 $ & $     0.12   \pm     0.02 $ &      2.210   &      0.002 \\  
 EXO~0422$-$086/RBS0540&     42   & $  298  \pm    59 $ & $     0.15  \pm      0.09 $ & $     0.89   \pm     0.04 $ &      2.070   &      0.004 \\  
    A3266              &    559   & $ 1174  \pm    41 $ & $     9.15  \pm      1.34 $ & $     8.33   \pm     0.34 $ &      3.133   &      0.017 \\  
    A0496              &    360   & $  687  \pm    28 $ & $     1.71  \pm      0.27 $ & $     2.56   \pm     0.12 $ &      2.001   &      0.004 \\  
    A3376              &    165   & $  798  \pm    46 $ & $     2.67  \pm      0.54 $ & $     1.28   \pm     0.09 $ &      5.341   &      0.989 \\  
    A3391              &     71   & $  716  \pm    62 $ & $     1.92  \pm      0.54 $ & $     2.35   \pm     0.12 $ &      3.801   &      0.043 \\  
    A3395s             &    215   & $  841  \pm    39 $ & $     3.14  \pm      0.54 $ & $     1.74   \pm     0.13 $ &      4.909   &      0.533 \\  
    A0576              &    237   & $  837  \pm    39 $ & $     3.12  \pm      0.54 $ & $     1.10   \pm     0.15 $ &      2.056   &      0.040 \\  
    A0754              &    470   & $  928  \pm    34 $ & $     4.31  \pm      0.65 $ & $     6.30   \pm     0.31 $ &      2.426   &      0.644 \\  
    A0780/Hydra~A      &     37   & $  687  \pm    82 $ & $     1.69  \pm      0.63 $ & $     2.94   \pm     0.14 $ &      1.642   &      0.012 \\  
    A1060              &    389   & $  652  \pm    21 $ & $     1.48  \pm      0.21 $ & $     0.32   \pm     0.05 $ &      1.932   &      0.001 \\  
    A1367              &    343   & $  639  \pm    24 $ & $     1.38  \pm      0.21 $ & $     1.01   \pm     0.05 $ &      4.425   &      0.552 \\  
    MKW4               &    145   & $  417  \pm    37 $ & $     0.40  \pm      0.11 $ & $     0.18   \pm     0.02 $ &      2.107   &      0.007 \\  
    ZwCl~1215.1$+$0400 &    154   & $  889  \pm    51 $ & $     3.70  \pm      0.74 $ & $     5.61   \pm     0.26 $ &      2.872   &      0.015 \\  
    NGC4636            &    115   & $  224  \pm    12 $ & $     0.07  \pm      0.01 $ & $     0.00   \pm     0.00 $ &      1.373   &      0.005 \\  
    A3526/Centaurus    &    235   & $  486  \pm    24 $ & $     0.61  \pm      0.11 $ & $     0.68   \pm     0.09 $ &      2.168   &      0.012 \\  
    A1644              &    307   & $  980  \pm    48 $ & $     5.13  \pm      0.92 $ & $     2.99   \pm     0.29 $ &      3.841   &      0.021 \\  
    A1650              &    220   & $  794  \pm    43 $ & $     2.58  \pm      0.50 $ & $     5.30   \pm     0.21 $ &      2.019   &      0.019 \\  
    A1651              &    222   & $  896  \pm    36 $ & $     3.77  \pm      0.60 $ & $     6.51   \pm     0.46 $ &      1.988   &      0.003 \\  
    A1656/Coma         &    972   & $  970  \pm    22 $ & $     5.05  \pm      0.62 $ & $     8.12   \pm     0.56 $ &      3.113   &      0.051 \\  
    NGC5044            &    156   & $  308  \pm    20 $ & $     0.17  \pm      0.04 $ & $     0.06   \pm     0.00 $ &      1.563   &      0.002 \\  
    A1736              &    148   & $  832  \pm    43 $ & $     3.05  \pm      0.57 $ & $     1.95   \pm     0.52 $ &      5.738   &      0.631 \\  
    A3558              &    509   & $  902  \pm    27 $ & $     3.94  \pm      0.53 $ & $     5.48   \pm     0.15 $ &      2.787   &      0.042 \\  
    A3562              &    265   & $ 1029  \pm    41 $ & $     6.01  \pm      0.94 $ & $     2.96   \pm     0.29 $ &      2.749   &      0.018 \\  
    A3571              &    172   & $  853  \pm    45 $ & $     3.32  \pm      0.63 $ & $     7.38   \pm     0.25 $ &      2.198   &      0.012 \\  
    A1795              &    179   & $  791  \pm    41 $ & $     2.58  \pm      0.48 $ & $     6.09   \pm     0.17 $ &      1.639   &      0.005 \\  
    A3581              &     83   & $  439  \pm    41 $ & $     0.45  \pm      0.14 $ & $     0.20   \pm     0.02 $ &      1.617   &      0.018 \\  
    MKW8               &    183   & $  450  \pm    25 $ & $     0.49  \pm      0.10 $ & $     0.37   \pm     0.04 $ &      4.175   &      0.024 \\  
RXC~J1504.1$-$0248/RBS1460&  208   & $  888  \pm    47 $ & $     3.38  \pm      0.64 $ & $    14.20   \pm     0.95 $ &      1.162   &      0.025 \\  
    A2029              &    202   & $ 1247  \pm    61 $ & $    11.00  \pm      1.97 $ & $     9.55   \pm     0.61 $ &      1.451   &      0.008 \\  
    A2052              &    168   & $  590  \pm    35 $ & $     1.08  \pm      0.22 $ & $     1.12   \pm     0.05 $ &      1.683   &      0.006 \\  
    MKW3S/WBL564       &     94   & $  599  \pm    42 $ & $     1.13  \pm      0.27 $ & $     1.66   \pm     0.07 $ &      1.644   &      0.038 \\  
    A2065              &    204   & $ 1146  \pm    47 $ & $     8.38  \pm      1.34 $ & $     5.19   \pm     0.39 $ &      2.134   &      0.039 \\  
    A2063              &    224   & $  646  \pm    33 $ & $     1.42  \pm      0.26 $ & $     1.36   \pm     0.06 $ &      2.259   &      0.010 \\  
    A2142              &    233   & $ 1008  \pm    46 $ & $     5.48  \pm      0.94 $ & $     9.69   \pm     0.97 $ &      2.360   &      0.004 \\  
    A2147              &    397   & $  859  \pm    32 $ & $     3.40  \pm      0.52 $ & $     3.21   \pm     0.47 $ &      4.296   &      0.009 \\  
    A2163              &    311   & $ 1498  \pm    61 $ & $    18.50  \pm      2.95 $ & $    63.78   \pm     5.24 $ &      2.518   &      0.074 \\  
    A2199              &    374   & $  733  \pm    29 $ & $     2.08  \pm      0.33 $ & $     2.02   \pm     0.14 $ &      1.758   &      0.012 \\  
    A2204              &    111   & $  917  \pm    99 $ & $     3.90  \pm      1.33 $ & $    13.94   \pm     0.76 $ &      1.384   &      0.006 \\  
    A2244              &    106   & $ 1116  \pm    63 $ & $     7.57  \pm      1.51 $ & $     5.73   \pm     1.09 $ &      1.611   &      0.014 \\  
    A2256              &    296   & $ 1216  \pm    45 $ & $    10.24  \pm      1.53 $ & $     8.42   \pm     0.39 $ &      2.683   &      0.104 \\  
    A2255              &    189   & $  998  \pm    55 $ & $     5.33  \pm      1.04 $ & $     5.98   \pm     0.48 $ &      4.142   &      0.213 \\  
    A3667              &    580   & $ 1073  \pm    37 $ & $     6.85  \pm      0.99 $ & $     8.75   \pm     0.20 $ &      3.361   &      0.100 \\  
AS1101/S\'{e}rsic~159$-$03 &     20   & $  422  \pm    55 $ & $     0.40  \pm      0.17 $ & $     1.37   \pm     0.06 $ &      1.627   &      0.009 \\  
    A2589              &     94   & $  762  \pm    57 $ & $     2.32  \pm      0.58 $ & $     1.14   \pm     0.07 $ &      1.969   &      0.008 \\  
    A2597              &     44   & $  525  \pm    54 $ & $     0.74  \pm      0.24 $ & $     2.75   \pm     0.11 $ &      1.533   &      0.011 \\  
    A2634              &    192   & $  721  \pm    38 $ & $     1.98  \pm      0.37 $ & $     0.74   \pm     0.07 $ &      4.825   &      0.003 \\  
    A2657              &     64   & $  764  \pm    92 $ & $     2.34  \pm      0.88 $ & $     1.54   \pm     0.12 $ &      2.325   &      0.018 \\  
    A4038              &    202   & $  764  \pm    37 $ & $     2.36  \pm      0.42 $ & $     0.99   \pm     0.08 $ &      1.822   &      0.010 \\  
    A4059              &    188   & $  674  \pm    43 $ & $     1.60  \pm      0.35 $ & $     1.70   \pm     0.09 $ &      1.910   &      0.014 \\  
    \hline 
    \hline
\end{tabular}
\label{t:log}
\end{center}
\hspace*{0.0cm}{\footnotesize 
}
\end{table*}

\begin{table*}
\begin{center}
  \caption[]{Ratios of X-ray bolometric luminosity measurements in
    different annuli for the whole HIFLUGCS sample of 64 clusters.}
\begin{tabular}{l|r|r|r|rr}
  \hline
  \hline
  Luminosity ratios                                     & Mean & Maximum& Minimum & Stddev  & Median\\
  \hline
  $L(0.2r_{500}\le R\le r_{500})/L(0.2r_{500}\le R\le 2.5r_{500})$        & 0.749   & 0.950  & 0.529   & 0.093 & 0.760 \\
  $L(R\le r_{500})/L(R\le 2.5r_{500})$                     & 0.842   & 0.968  & 0.594   & 0.088 & 0.863 \\
  $L(0.2r_{500}\le R\le r_{500})/L(R\le r_{500})$          & 0.543   & 0.826  & 0.140   & 0.158 & 0.539 \\
  $L(0.2r_{500}\le R\le 2.5r_{500})/L(R\le 2.5 r_{500})$   & 0.606   & 0.886  & 0.189   & 0.161 & 0.615 \\
  \hline 
  \hline
\end{tabular}
\label{t:lbratio}
\end{center}
\hspace*{0.0cm}{\footnotesize 
}
\end{table*}

\begin{table*}
\begin{center}
  \caption[]{Gaussian fit parameters of the histograms of the
    logarithmic values of $c_{\rm L}$ and ${\Delta}R/r_{500}$ for
    the full HIFLUGCS sample of all 64 clusters.}
\begin{tabular}{l|r|r}
  \hline
  \hline
  Distribution    & Mean & FWHM  \\
  \hline
  $\log_{10}c_{\rm L}$    &  $0.31\pm 0.03$ & $0.13\pm 0.03$  \\
  \hline
  $\log_{10}({\Delta}R/r_{500})$  &  $-1.91 \pm 0.04$& $0.45\pm 0.04$  \\
    \hline 
  \hline
\end{tabular}
\label{t:histcloff}
\end{center}
\hspace*{0.0cm}{\footnotesize 
}
\end{table*}

\begin{table*}
\begin{center}
  \caption[]{Best power-law fits,
    $\log_{10}\widetilde{L}=A\log_{10}\widetilde{M}+B$, in which
    $\widetilde{L}=\frac{L^{\rm cor}}{E(z)10^{44}{\rm erg/s}}$ and
    $\widetilde{M}=\frac{M\,E(z)}{10^{14}M_{\odot}}$, of the $L^{\rm
      cor}-M$ relations. Note that all disturbed clusters as well as
    NCC clusters have at least 45 member galaxies with spectroscopic
    redshifts per cluster in the current observational sample.}
\begin{tabular}{l|lllll}
  \hline
  \hline
Sample & Number      & $B$            & $A$                 & $\sigma_{\log_{10}L,{\rm intrinsic}}$ & $B$ for $A=4/3$\\
       & of clusters &                &                     &                                &                 \\
    \hline
Whole        &  63   & $ -0.16\pm    0.06$ & $ 1.29\pm   0.09$ & $ 0.33\pm   0.04$ & $-0.18 \pm  0.04 $\\ 
 Undisturbed &  39   & $ -0.14\pm    0.07$ & $ 1.36\pm   0.13$ & $ 0.39 \pm  0.05$ & $-0.12 \pm  0.06 $\\ 
   Disturbed &  24   & $ -0.29\pm    0.10$ & $ 1.26\pm   0.14$ & $ 0.22 \pm  0.03$ & $-0.28 \pm  0.05 $\\ 
    CC      &  27   & $  +0.02\pm    0.08$ & $ 1.45\pm   0.16$ & $ 0.41 \pm  0.07$ & $-0.04 \pm  0.08 $\\ 
   NCC      &  36   & $  -0.27\pm    0.09$ & $ 1.41\pm   0.13$ & $ 0.23 \pm  0.02$ & $-0.28 \pm  0.04 $\\ 
\hline
 Whole ($n_{\rm gal}\ge45$)       &  57   & $ -0.35\pm    0.05$ & $ 1.44\pm   0.08$ & $ 0.27 \pm  0.02$ & $-0.24 \pm  0.04 $\\ 
 Undisturbed ($n_{\rm gal}\ge45$) &  33   & $ -0.28\pm    0.06$ & $ 1.57\pm   0.11$ & $ 0.29 \pm  0.03$ & $-0.21 \pm  0.06 $\\
  CC ($n_{\rm gal}\ge45$)         &  21   & $ -0.16\pm    0.07$ & $ 1.66\pm   0.13$ & $ 0.29 \pm  0.06$ & $-0.16 \pm  0.08 $\\
  \hline
  \hline
\end{tabular}
\label{t:lm}
\end{center}
\hspace*{0.0cm}{\footnotesize 
}
\end{table*}

\clearpage

\appendix

\section{Impact of flux loss of the aperture magnitude}
\label{a:fluxloss}

\subsection{Fraction of recovered flux}
\label{a:fracflux}

We used a S\'{e}rsic profile for the two-dimensional (2D) surface
brightness profiles of galaxies (e.g. Peng et al. 2002), $ \Sigma(x,y)
\propto \exp \left[ -\kappa \left(\frac{r_{\rm ell}}{R_{\rm e}}
  \right)^{1/n}-1 \right]$ with $ r_{\rm ell} = \left[ \left\vert x
  \right\vert ^{(c+2)}+\left\vert \frac{y}{q} \right\vert ^{(c+2)}
  \right]^{1/(c+2)}$. The parameter $q$ is the ellipticity, and we
used its distribution observed in SDSS (Fig.~3 in Hyde \& Bernardi
2009, peaked at $\sim$0.8). For a de Vaucouleurs profile, $c=0$,
$n=4$, $\kappa=7.67$ for $n=4$, and $\Sigma(x,y) \propto \exp \left[
  -7.67 \left(r_{\rm ell}/R_{\rm e}\right)^{1/4}\right]\; $. Following
Bernardi et al. (2007), we computed the effective radius in units of
kpc from the $r$-band absolute magnitude for a passive, elliptical
galaxy, $\langle \log R_{\rm e} \vert M_r \rangle = -0.681
(M_r+21)/2.5 + 0.343$, using the spectral energy distribution template
in Maraston et al. (2009).

\begin{figure*}
\begin{center}
\hspace{-1cm}
\includegraphics[width=9.5cm]
{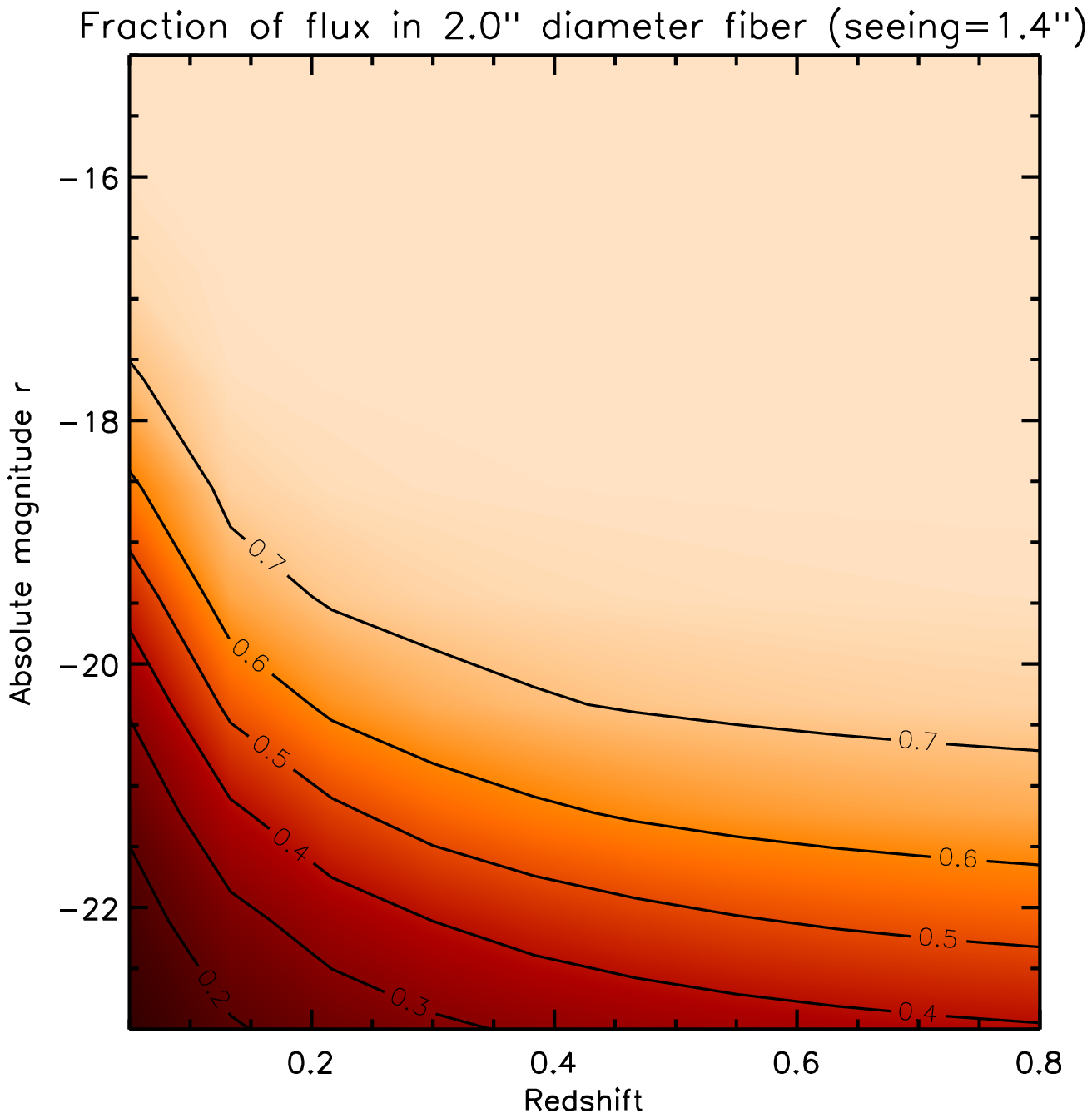}
\hspace{-1cm}
\includegraphics[width=9.5cm]
{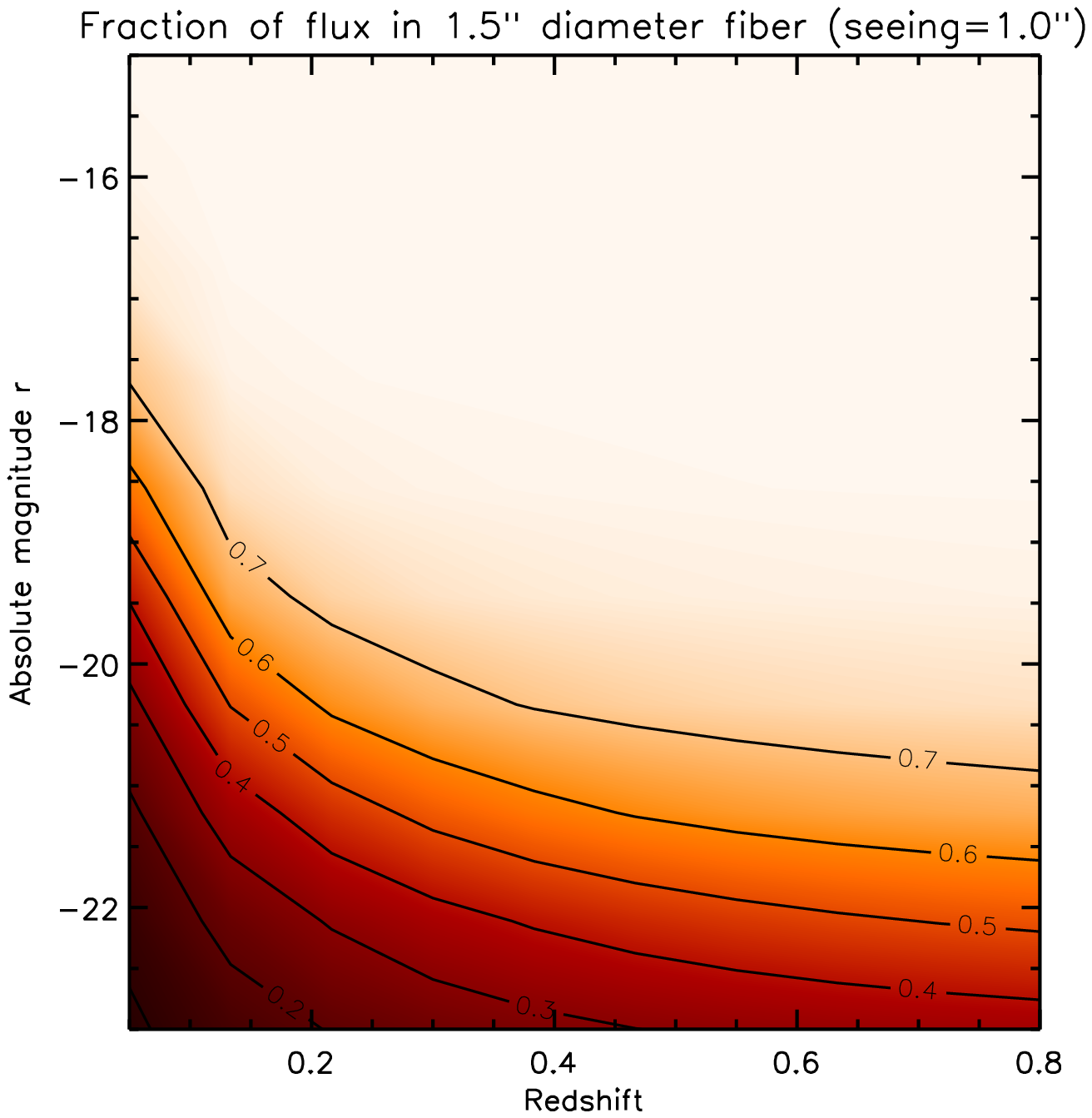} 
\end{center}
\caption{Fractions of recovered flux $f_{\rm synth}$ in the fibers of the
SPIDERS (left panel)
and 4MOST (right panel)
configurations. The $x$-axis and $y$-axis are the galaxy redshift and
$r$-band absolute magnitude, respectively.}
\label{f:fracflux}
\end{figure*}

We assume $1\farcs4$ seeing for SPIDERS and $1\arcsec$ seeing for
4MOST (private communication with T. Dwelly), in which the seeing is
better than the given value for 90\% of the time.
The diameter of the aperture is $\varnothing = 2\arcsec$ for SPIDERS
and $1\farcs5$ for 4MOST. The resulting flux is computed by
integrating the surface brightness profile convolved with the seeing
within the aperture. The fractions of recovered flux $f_{\rm synth}$
in the fibers of the SPIDERS and 4MOST surveys are shown in
Fig.~\ref{f:fracflux}. Our results are consistent with that of the
objects classified as ``LRG'' or ``GALAXY'' which have spectroscopic
redshifts from the BOSS in SDSS DR11.

\subsection{Impact on our sampling}

For galaxies with an intrinsic magnitude only slightly brighter than
the limiting magnitude cut, its aperture magnitude may become fainter
than the limiting magnitude when considering flux loss.  We inspected
this impact on our sampling by comparing the aperture magnitudes of
the member galaxies used in the re-sampling with the limiting
magnitude cut. Although there is quite a dramatic difference
($\sim$1--1.5 magnitudes) between the total magnitude and the
$2\arcsec$ aperture magnitude, we found that the aperture magnitudes
of all member galaxies are well above the limiting magnitude cut for
the clusters at $z_{\rm in}<0.4$ and $z_{\rm in}=0.8$. This is because
the galaxies are much brighter than the limiting magnitude at the low
redshift of $z_{\rm in}=0.2$, and the angular sizes of the galaxies
are well within the aperture size at the high redshift of $z_{\rm
  in}=0.8$ such that flux loss plays no role. At redshifts of $z_{\rm
  in}=$0.4 and 0.6, the aperture magnitude considering flux loss leads
to a number of member galaxies fainter than the limiting magnitude in
a couple of clusters among 63 clusters in the re-sampling. No more
than three out of 63 clusters are affected by the impact of flux
loss. We thus consider this effect negligible. Note that the flux loss
plays a major role in degrading the spectral quality, which is not
part of this study.

\end{document}